\pdfoutput=1
\documentclass[12pt,a4paper,twoside,fleqn]{article}
\usepackage{color,graphicx,lscape,here,geometry,setspace,multirow,enumitem}
\usepackage[dvipsnames]{xcolor}
\usepackage{graphicx}
\usepackage[longnamesfirst]{natbib}

\usepackage{authblk}
\usepackage{silence}

\usepackage{rotating}
\usepackage{multirow}

\usepackage{booktabs}
\usepackage{diagbox}
\usepackage{color}
\usepackage{setspace}
\usepackage{algorithm2e}
\usepackage{enumitem}
\usepackage{tikz}

\usepackage{dcolumn} 
\newcolumntype{d}[1]{D{.}{.}{#1}}   

\definecolor{nblue}{HTML}{000660}
\usepackage[colorlinks=true,urlcolor=nblue,linkcolor=nblue,citecolor=nblue]{hyperref}
\usepackage{bigstrut}

\usepackage{tabularx}
\usepackage{threeparttable}
\usepackage{multirow}

\usepackage{etoolbox} 
\makeatletter
\patchcmd{\BR@backref}{\newblock}{\newblock[}{}{}
\patchcmd{\BR@backref}{\par}{]\par}{}{}
\makeatother

\usepackage{pdflscape}
\usepackage{afterpage}

\usepackage{longtable}
\usepackage{array}
\newcolumntype{C}[1]{>{\centering\arraybackslash}p{#1}}

\usepackage{amsmath}
\usepackage{amssymb}

\usepackage{authblk}
\usepackage{silence}

\usepackage[hang,flushmargin]{footmisc}

\usepackage[hang]{footmisc}
\usepackage[british]{babel}

\pdfoutput=1
\usepackage[title,titletoc]{appendix}
\makeatletter

\renewenvironment{appendices}{%
    \begin{oldappendices}%
    \renewcommand{\thefigure}{\ifnum \c@section>\z@ \thesection.\fi\@arabic\c@figure}%
    \@addtoreset{figure}{section}%
    \renewcommand{\thetable}{\ifnum \c@section>\z@ \thesection.\fi\@arabic\c@table}%
    \@addtoreset{table}{section}}{%
    \end{oldappendices}%
}\makeatother

\usepackage{titlesec} 
\titleformat{\section}[block]{\bfseries\sffamily\large}{\thesection. }{0em}{\MakeUppercase} 
\titleformat{\subsection}[block]{\bfseries\sffamily\large}{\thesubsection. }{0em}{} 
\titleformat{\subsubsection}[block]{\large}{}{0em}{\itshape} 

\let\natbibcitet\citet
\renewcommand\citet{\bibpunct{(}{)}{,}{a}{,}{,}\natbibcitet}
\let\natbibcitep\citep
\renewcommand\citep{\bibpunct{(}{)}{;}{a}{,}{;}\natbibcitep}
\newcommand{\bi}{\begin{itemize}}
\newcommand{\ei}{\end{itemize}}
\newcommand{\be}{\begin{equation}}
\newcommand{\ee}{\end{equation}}
\defcitealias{ieo14}{IEO, 2014}

\long\def\symbolfootnote[#1]#2{\begingroup%
\def\thefootnote{\fnsymbol{footnote}}\footnote[#1]{#2}\endgroup}

\makeatletter
\def\ubar#1{\underline{\sbox\tw@{$#1$}\dp\tw@\z@\box\tw@}}
\def\obar#1{\overline{\sbox\tw@{$#1$}\dp\tw@\z@\box\tw@}}
\makeatother

\usepackage{bm}
\usepackage{caption}
\usepackage{subcaption}

\makeatletter\let\p@subfigure\thefigure\makeatother

\floatstyle{plaintop}
\restylefloat{table}

\usepackage{fancyhdr} 
\newcommand{\diag}{\text{diag}}
\newcommand{\EA}{\text{EA}}
\newcommand{\US}{\text{US}}

\newcommand{\MP}{\text{MP}}
\newcommand{\CI}{\text{CI}}

\usepackage{cleveref}
\crefname{chapter}{Chapter}{Chapters}
\crefname{section}{Section}{Sections}
\crefname{subsection}{Section}{Sections}
\crefname{subsubsection}{Section}{Sections}
\crefname{figure}{Figure}{Figures}
\crefname{table}{Table}{Tables}
\crefname{equation}{Equation}{Equations}
\crefname{appendix}{Appendix}{Appendices}
\crefname{appendices}{Appendix}{Appendices}

\crefname{appsec}{Appendix}{Appendices}

\makeatletter

\def\Autoref#1{%
  \begingroup
  \edef\reserved@a{\cpttrimspaces{#1}}%
  \ifcsndefTF{r@#1}{%
    \xaftercsname{\expandafter\testreftype\@fourthoffive}
      {r@\reserved@a}.\\{#1}%
  }{%
    \ref{#1}%
  }%
  \endgroup
}
\def\testreftype#1.#2\\#3{%
  \ifcsndefTF{#1autorefname}{%
    \def\reserved@a##1##2\@nil{%
      \uppercase{\def\ref@name{##1}}%
      \csn@edef{#1autorefname}{\ref@name##2}%
      \autoref{#3}%
    }%
    \reserved@a#1\@nil
  }{%
    \autoref{#3}%
  }%
}
\makeatother
\def\titletext{High-frequency and heteroskedasticity identification in multicountry models: Revisiting spillovers of monetary shocks}

\title{\sffamily\huge{\textbf{\titletext}}}
\author{}
\date{}


\def\equationautorefname~#1\null{%
  Eq.~(#1)\null
}
\def\equationautorefname~#1\null{
Eq.~(#1)\null
}

\usepackage[utf8, latin1]{inputenc}                 

\begin{document}

\maketitle
\vspace*{-6.5em}
\vspace*{2.5em}
\normalsize
\begin{center}
\begin{minipage}{.49\textwidth}
  \large\centering Michael \MakeUppercase{Pfarrhofer}\\[0.25em]
  \normalsize WU Vienna
\end{minipage}
\begin{minipage}{.49\textwidth}
  \large\centering Anna \MakeUppercase{Stelzer}\\[0.25em]
  \normalsize Oesterreichische Nationalbank
\end{minipage}
\end{center}

\vspace*{1em}
\doublespacing
\begin{center}
\begin{minipage}{0.9\textwidth}
\noindent\small 
We explore the international transmission of monetary policy and central bank information shocks originating from the United States and the euro area. Employing a panel vector autoregression, we use macroeconomic and financial variables across several major economies to address both static and dynamic spillovers. To identify structural shocks, we introduce a novel approach that combines external instruments with heteroskedasticity-based identification and sign restrictions. Our results suggest significant spillovers from European Central Bank and Federal Reserve policies to each other's economies, global aggregates, and other countries. These effects are more pronounced for central bank information shocks than for pure monetary policy shocks, and the dominance of the US in the global economy is reflected in our findings.

{\sffamily\textbf{JEL}}: C32, E52, E58, F42\\
{\sffamily\textbf{KEYWORDS}}: Bayesian panel vector autoregression, structural identification, external instruments, factor model, central bank information
\end{minipage}
\end{center}

\singlespacing\vfill\noindent{\footnotesize\textit{Contact}: Anna Stelzer. Monetary Policy Section, Oesterreichische Nationalbank. \textit{Address}: Otto-Wagner-Platz 3, 1090 Vienna, Austria. \textit{Email}: \href{mailto:anna.stelzer@oenb.at}{anna.stelzer@oenb.at}. We thank Niko Hauzenberger, Florian Huber and participants of the 27th ICMAIF for valuable comments and suggestions. Karen Spisso provided excellent research assistance. Stelzer acknowledges funding by the Oesterreichische Nationalbank (project no. $18763$) during previous employment. This paper is a substantially revised version of ``The international effects of central bank information shocks.''
\textit{Disclaimer:} The views expressed in this paper do not necessarily reflect those of the Oesterreichische Nationalbank or the Eurosystem.}

\thispagestyle{empty}

\newpage







\doublespacing
\section{Introduction}\label{sec:intro}

In an interconnected global financial system, understanding spillovers of monetary policy and central bank communication has become crucial for policymakers and market participants alike. While the international transmission of monetary policy has been extensively studied, less is known about the cross-border effects of central bank information shocks --- reactions of (market) expectations to the information content of central bank communication beyond changes in policy rates \citep[see, e.g.,][]{nakamura2018high,jarocinski2020deconstructing,gardner2022words,bauer2023alternative}. Our paper adds to this literature by investigating and re-examining the international spillovers of central bank information (CI) shocks alongside monetary policy (MP) shocks, focusing on the dynamics between the European Central Bank (ECB) and the Federal Reserve Bank (Fed).

Given the importance of the US dollar as a global currency and the impact of the Fed's monetary policy actions on the global financial cycle, a large body of literature examines the spillover effects of the Fed's policy decisions on the global economy \citep[see, e.g.,][]{kim2001international,ehrmann2011stocks,Eickmeier2011,miranda2020us,dedola2017if,gerko2017monetary,crespo2019spillovers,degasperi2020global,georgiadis2023global}. The international effects of policy shocks caused by central banks other than the Fed have received less attention, and the literature on the international transmission of ECB policy actions is limited \citep[for exceptions, see][]{ehrmann2011stocks,potjagailo2017spillover,feldkircher2020international,miranda2022tale}. 

Related to the work of \citet{jarocinski2022central}, who assesses a restricted variant of spillovers of euro area (EA) shocks to the US economy and vice versa, our analysis extends to a joint model of the dynamic evolution of these two economies, alongside several others, namely, the United Kingdom (UK), Japan (JP) and Canada (CA). Thus, the data include the G7 economies but are based on aggregate EA data rather than including its member states individually. Our work is also closely related to \citet{ca2023making}, who study a spillover effect of pure monetary policy shocks (purged of information effects) using separately estimated and identified, economy-specific, VAR models. In contrast, one of the contributions of this paper is to provide a unified multicountry framework for identification, estimation, and inference in order to study possible spillover from both monetary policy and information spillovers.

Methodologically, we introduce several innovations to the existing literature. We combine a Bayesian panel vector autoregression (PVAR), similar to that of \citet{huber2018bayesian}, with a model for measurement errors of monetary policy shock instruments, inspired by \citet{caldara2019monetary}. External instruments constructed from high-frequency changes in relevant asset prices surrounding policy announcement dates are nowadays routinely employed in empirical macroeconomics to identify (monetary) policy shocks \citep[see][for a recent review]{bauer2023reassessment}. 

We include these instruments in our PVAR and use a factor stochastic volatility (FSV) model on the reduced form errors of the resulting multivariate system. This allows us to capture multiple structural shocks across multiple economies simultaneously. Specifically, our implementation relies on a mix of zero and sign restrictions, imposed with truncated normal priors on the factor loadings. It achieves identification of the shocks in conjunction with the assumption of static heteroskedastic factors \citep[see also][]{korobilis2022new,gambetti2023agreed,chan2022large}. As such, our paper is related to the recent literature that combines identification schemes in a unified framework \citep[see, e.g.,][]{schlaak2023monetary,carriero2024blended}.\footnote{The proposed approach explicitly introduces (non-data) identifying information via a prior, similar in spirit to \citet{baumeister2015sign}. Our paper also relates to the literature that exploits heteroskedasticity for identification purposes \citep[see, among many others,][]{sentana2001identification,rigobon2003identification,gurkaynak2020missing,lewis2021identifying,bertsche2022identification,griller2024financial}, providing an alternative to the rotation-based sign restrictions used in previous studies such as \citet{jarocinski2020deconstructing}, \citet{jarocinski2022central} or \citet{ca2023making}.}

This modeling approach is designed to allow a comprehensive analysis of the cross-border effects of both monetary policy and central bank information shocks, and the main features can be summarized as follows. First, the FSV model combined with sign and zero restrictions on the loadings matrix allows for point identification of the structural shocks (which we treat as latent quantities) from a set of external instruments, in a unified framework. It is worth noting that our approach does not require the shocks to be invertible. Second, the PVAR allows us to simultaneously investigate static (contemporaneous) spillovers of shocks originating in a single economy, and the dynamic (higher-order) propagation of these shocks which involves the domestic transmission as well as spillovers. These latter aspects are crucial to studying dynamic interdependencies between the respective domestic and foreign economies, including also spillback effects \citep[see, e.g.,][]{breitenlechner2022goes}.

Our findings have important implications for the conduct of monetary policy in an economically integrated global economy. The empirical results indicate that monetary policy and central bank information shocks from both the Fed and the ECB generate significant spillovers (both static and dynamic) to each other and to other major economies, confirming previous empirical evidence. On average, we find that central bank information shocks generate slightly stronger spillovers than pure monetary policy shocks. Relatedly, the spillovers from US monetary policy shocks tend to be somewhat stronger and more persistent, reflecting the dominance of the US economy. The analysis also shows that domestic and foreign impacts differ considerably, with cross-border effects being more pronounced in financial markets than in real economic variables such as industrial production or consumer prices.

The remainder of the paper is organized as follows. Section \ref{sec:econometrics} proposes the multicountry econometric framework and discusses structural identification. Section \ref{sec:data} provides further information on the dataset and model specification, while Section \ref{sec:results} presents our empirical findings. Section \ref{sec:conclusion} concludes.

\section{Econometric framework}\label{sec:econometrics}
\subsection{A multicountry vector autoregression}
Our multicountry framework represents a variant of an unrestricted PVAR and is set up in its most general form in this sub-section \citep[see][for a recent review of multicountry models]{feldkircher2020factor}. We subsequently tailor several aspects to our application, especially our approach to modeling the reduced form errors. Let $\bm{\mathrm{y}}_{i,t}$ be an $M_i\times1$-vector of macroeconomic and financial variables for countries $i = 1,\hdots, N,$ at time $t = 1,\hdots,T$. Suppose that we observe $S_i$ instruments or proxies $\bm{m}_{i,t}$ for structural shocks in a subset of countries $i = 1,\hdots,\tilde{N},$ where $\tilde{N} < N$ and $i = 1,\hdots,\tilde{N},\tilde{N}+1,\hdots,N$. 

Define $\bm{y}_t = (\bm{m}_{1,t}',\hdots,\bm{m}_{\tilde{N},t}',\bm{\mathrm{y}}_{1,t}',\hdots,\bm{\mathrm{y}}_{N,t}')'$ of size $M\times1$, where $M = \sum_{j=1}^{\tilde{N}}S_j + \sum_{i=1}^N M_i$. We assume that this vector follows a VAR process:
\begin{equation}
    \bm{y}_t = \bm{A}_1 \bm{y}_{t-1} + \hdots + \bm{A}_P \bm{y}_{t-P} + \bm{\epsilon}_t,\label{eq:pVAR}
\end{equation}
which establishes a PVAR with the reduced form errors stacked in $\bm{\epsilon}_t$. For notational simplicity, we omit any deterministic terms here but note that all of our implementations feature an intercept. The $M\times M$ matrices $\bm{A}_p$, for $p = 1,\hdots, P,$ feature a block structure, with sub-matrices modeling the domestic economy and dynamic cross-country spillovers, respectively:
\begin{equation*}
\left[\begin{array}{c}
\bm{m}_{1,t}\\
\vdots\\
\bm{m}_{\tilde{N},t}\\[0.5ex]
\hline
\bm{\mathrm{y}}_{1,t}\\
\vdots\\
\bm{\mathrm{y}}_{N,t}
\end{array}\right] = \sum_{p=1}^P \left[
\begin{array}{ccc|ccc}
\bm{A}_{p[11]}^{(mm)} & \hdots & \bm{A}_{p[1\tilde{N}]}^{(mm)} & \bm{A}_{p[11]}^{(my)} & \hdots & \bm{A}_{p[1N]}^{(my)} \\
\vdots & \ddots & \vdots & \vdots & \ddots & \vdots\\
\bm{A}_{p[\tilde{N}1]}^{(mm)} & \hdots & \bm{A}_{p[\tilde{N}\tilde{N}]}^{(mm)} & \bm{A}_{p[\tilde{N}1]}^{(my)} & \hdots & \bm{A}_{p[\tilde{N}N]}^{(my)} \\[0.5ex]
\hline
\bm{A}_{p[11]}^{(ym)} & \hdots & \bm{A}_{p[1\tilde{N}]}^{(ym)} & \bm{A}_{p[11]}^{(yy)} & \hdots & \bm{A}_{p[1N]}^{(yy)} \\
\vdots & \ddots & \vdots & \vdots & \ddots & \vdots\\
\bm{A}_{p[N1]}^{(ym)} & \hdots & \bm{A}_{p[N\tilde{N}]}^{(ym)} & \bm{A}_{p[N1]}^{(yy)} & \hdots & \bm{A}_{p[NN]}^{(yy)}
\end{array}\right]
\left[\begin{array}{c}
\bm{m}_{1,t-p}\\
\vdots\\
\bm{m}_{\tilde{N},t-p}\\[0.5ex]
\hline
\bm{\mathrm{y}}_{1,t-p}\\
\vdots\\
\bm{\mathrm{y}}_{N,t-p}
\end{array}\right] + \left[
\begin{array}{c}
\bm{\epsilon}_{1,t}^{(m)}\\
\vdots\\
\bm{\epsilon}_{\tilde{N},t}^{(m)}\\[0.5ex]
\hline
\bm{\epsilon}_{1,t}^{(y)}\\
\vdots\\
\bm{\epsilon}_{N,t}^{(y)}
\end{array}\right].
\end{equation*}

More specifically, the matrices $\bm{A}_{p[ij]}^{(mm)}, \bm{A}_{p[ij]}^{(my)}, \bm{A}_{p[ij]}^{(ym)}$ capture the dynamics of the instruments and linkages between the instruments and all other variables in the system of equations. The matrices $\bm{A}_{p[ii]}^{(yy)}$ relate to domestic dynamics, while $\bm{A}_{p[ij]}^{(yy)}$, for $i\neq j$, measure the dynamic relationships across economies. Analogously, for second moments, we note that Var$(\bm{\epsilon}_{i,t}^{(y)}) = \mathbb{E}(\bm{\epsilon}_{i,t}^{(y)}\bm{\epsilon}_{i,t}^{(y)}{'})$ measures the contemporaneous relationships across domestic variables within country $i$. By contrast, Cov$(\bm{\epsilon}_{i,t}^{(y)},\bm{\epsilon}_{j,t}^{(y)}) = \mathbb{E}(\bm{\epsilon}_{i,t}^{(y)}\bm{\epsilon}_{j,t}^{(y)}{'})$ for $i\neq j$ captures static spillovers. 

To efficiently model the full covariance structure, we use a factor model on the reduced form errors $\bm{\epsilon}_t$:
\begin{equation}
\bm{\epsilon}_t = \bm{L} \bm{\mathrm{f}}_t + \bm{\eta}_t, \quad \bm{\mathrm{f}}_t \sim \mathcal{N}(\bm{0}_Q,\bm{H}_t), \quad \bm{\eta}_t \sim \mathcal{N}(\bm{0}_M,\bm{\Omega}), \label{eq:fsv}
\end{equation}
where $\bm{\mathrm{f}}_t$ is a set of $Q$ static latent factors with time-varying $Q\times Q$-covariance matrix $\bm{H}_t = \diag(\exp(h_{1,t}),\hdots,\exp(h_{Q,t}))$, $\bm{L}$ is an associated $M \times Q$ factor loadings matrix, and $\bm{\eta}_t$ of size $M$ are homoskedastic Gaussian measurement errors with diagonal $M\times M$-covariance matrix $\bm{\Omega} = \diag(\omega_{1}^2,\hdots,\omega_{M}^2)$. To implement our identification scheme later on, we focus on cases where $Q > (\sum_{j=1}^{\tilde{N}} S_j)$. We also make the standard assumption that the factors and measurement errors are uncorrelated, i.e., $\mathbb{E}(\bm{\mathrm{f}}_t\bm{\eta}_t') = \bm{0}$. This establishes a specific variant of FSV \citep[see][in a VAR context]{kastner2020sparse}. The representation using auxiliary variables in Eq. (\ref{eq:fsv}) implies a full covariance matrix of the form $\bm{\Xi}_t = \bm{L}\bm{H}_t\bm{L}' + \bm{\Omega}$. An important implication of this framework is thus that the covariances are driven purely by the latent factors. This reflects the notion that the number of ``primitive'' economic shocks, especially in large systems or multicountry models such as ours, is likely to be small \citep[see, e.g.,][]{bai2007determining,ramey2016macroeconomic}.

We make these assumptions about the reduced form errors for three reasons. First, to efficiently reduce the dimensionality of the potentially huge-dimensional covariance matrix of the PVAR. Second, we may exploit this setup to estimate the system equation-by-equation, which is beneficial from a computational viewpoint. Third, and arguably most importantly, we may use this framework to achieve identification of the structural shocks (that are potentially measured imperfectly by the instruments). This is equivalent to identifying the factor model, as we will illustrate below, which can be achieved by imposing a mix of sign and zero restrictions on the loadings matrix $\bm{L}$ (see \citet{korobilis2022new} and \citet{chan2022large} for a formal econometric discussion). Specifically, we split the vector of factors into two parts, $\bm{\mathrm{f}}_t = (\bm{f}_t',\tilde{\bm{f}}_t')'$ with $Q = Q_f + Q_{\tilde{f}}$, where $\bm{f}_t$ are $Q_f$ identified factors related to the instruments, and $\tilde{\bm{f}}_t$ are $Q_{\tilde{f}}$ additional factors. The former are used to identify the contemporaneous responses to structural shocks of interest, explicitly linked to the observed instruments. The latter serve as a device to model the remaining part of the cross-country covariance structure.

Our framework is rather general and capable of identifying both structural country-specific and common international shocks. For instance, we may use a version of our model when we want to use multiple instruments to identify multiple structural shocks (potentially in several economies, i.e., when $Q_f = \sum_{j = 1}^{\tilde{N}} S_j$), but also when there are multiple instruments for a smaller number of shocks (or a single shock), i.e., when $Q_f < \sum_{j = 1}^{\tilde{N}} S_j$, or when there generally is only a single shock of interest. That is, our general framework may for example also be applied in conventional VARs dealing with the single-economy case ($\tilde{N} = N = 1$).

However, in our empirical work, we focus on MP and CI shocks inspired by the framework of \citet{jarocinski2020deconstructing}. For illustration, we start with a multicountry version of their approach in the next sub-section. 

\subsection{Identifying structural factors in a multicountry setting}\label{subsec:identifyingstrufac}
To set the stage, suppose we observe surprises in interest rate futures ($I$) and stock markets ($S$) for the EA and the US (which is what we do in our application). That is, $\tilde{N} = 2$, and $\bm{m}_{\EA,t} = (m_{\EA,t}^{(I)},m_{\EA,t}^{(S)})'$ and $\bm{m}_{\US,t} = (m_{\US,t}^{(I)},m_{\US,t}^{(S)})'$ such that $S_\EA = S_\US = 2$, with corresponding economic and financial series $\bm{\mathrm{y}}_{\EA,t}$ and $\bm{\mathrm{y}}_{\US,t}$. In addition, we have $\bm{\mathrm{y}}_{\text{R},t} = (\bm{\mathrm{y}}_{3,t}',\hdots,\bm{\mathrm{y}}_{N,t}')'$, which collects the endogenous variables for all remaining (R) economies. 

We introduce the first set of (zero) restrictions needed for identification by writing Eq. (\ref{eq:pVAR}) as:
\begin{equation*}
\left[\begin{array}{c}
m_{\EA,t}^{(I)}\\
m_{\EA,t}^{(S)}\\
m_{\US,t}^{(I)}\\
m_{\US,t}^{(S)}\\[0.5ex]
\hline
\bm{\mathrm{y}}_{\EA,t}\\
\bm{\mathrm{y}}_{\US,t}\\
\bm{\mathrm{y}}_{\text{R},t}
\end{array}\right] = \sum_{p=1}^P \bm{A}_p \bm{y}_t +
\underbrace{\left[\begin{array}{cc|c}
\bm{L}_{\EA}^{(m)} & \bm{0} & \bm{0} \\
\bm{0} & \bm{L}_{\US}^{(m)} & \bm{0}\\[0.5ex]
\hline
\bm{L}_{\EA}^{(\EA,y)} & \bm{L}_{\EA}^{(\US,y)} & \tilde{\bm{L}}_{\EA}^{(y)}\\
\bm{L}_{\US}^{(\EA,y)} & \bm{L}_{\US}^{(\US,y)} & \tilde{\bm{L}}_{\US}^{(y)}\\
\bm{L}_{\text{R}}^{(\EA,y)} & \bm{L}_{\text{R}}^{(\US,y)} & \tilde{\bm{L}}_{\text{R}}^{(y)}\\
\end{array}\right]}_{\bm{L}}
\underbrace{\left[\begin{array}{c}
f_{\EA,t}^{(\MP)}\\
f_{\EA,t}^{(\CI)}\\
f_{\US,t}^{(\MP)}\\
f_{\US,t}^{(\CI)}\\[0.5ex]
\hline
\tilde{\bm{f}}_t
\end{array}\right]}_{\bm{\mathrm{f}}_t}
+
\left[\begin{array}{c}
{\eta}_{\EA,t}^{(I,m)}\\
{\eta}_{\EA,t}^{(S,m)}\\
{\eta}_{\US,t}^{(I,m)}\\
{\eta}_{\US,t}^{(S,m)}\\[0.5ex]
\hline
\bm{\eta}_{\EA,t}^{(y)}\\
\bm{\eta}_{\US,t}^{(y)}\\
\bm{\eta}_{\text{R},t}^{(y)}
\end{array}\right],
\end{equation*}
where identification of the MP and CI shocks for the EA and the US then boils down to identifying the upper (non-zero) blocks of the $Q_f = 4$ latent factors. In this regard, we assume several zero restrictions on the blocks of loadings that are associated with the instrument equations. Specifically, the instruments are uncorrelated across economies, and these surprises do not load on the remaining factors in $\tilde{\bm{f}}_t$ (which corresponds to an exclusion restriction that one would also encounter in the context of standard instrumental variables, IVs, or proxy VARs, see, e.g., \citealp{mertens2013dynamic}).\footnote{If economic and statistical identification of additional domestic or international shocks and/or factors is desired, one may do so by imposing further sign and/or zero restrictions on the loadings matrix $\bm{L}$, see \citet{korobilis2022new}. Apart from the restrictions we explicitly mention, we leave the loadings unrestricted because it is not necessary in the context of our application.} The upper right zero block in $\bm{L}$ thus rules out that other economic or financial shocks affect the instruments contemporaneously (i.e., they are assumed to be exogenous), and the off-diagonal zero blocks in the upper left block are needed to distinguish between the origin (EA or US) of the respective shock. In other words, these latter restrictions impose that the US shock is reflected only in the US instruments, and analogously for the EA, which is again very similar in spirit to a standard exclusion restriction in IV estimation.

To separate the MP and CI shocks within a given economy $i\in\{\EA,\US\}$, we need to impose --- in addition to the zero restrictions --- sign restrictions on the respective block of loadings $\bm{L}_i^{(m)}$. The loadings $\bm{L}_\EA^{(m)}$ and $\bm{L}_\US^{(m)}$ govern the static impact of the domestic shocks in the EA and US, respectively, on the corresponding domestic instruments. It is instructive to consider the link between the observed surprises, reduced form errors, and structural factors. The corresponding set of measurement equations is:\footnote{We note that this subset of equations is only used for illustrative purposes, and the factors, in general, may load on any of the vectors $\bm{\mathrm{y}}_{j,t}$ featured in Eq. (\ref{eq:pVAR}). Indeed, the corresponding loadings $\bm{L}_j^{(i,y)}$ measure impact responses, as we will show below.}
\begin{equation*}
\begin{bmatrix}
    m_{i,t}^{(I)}\\
    m_{i,t}^{(S)}
\end{bmatrix} = 
\underbrace{\begin{bmatrix}
+ & + \\
- & +
\end{bmatrix}}_{\bm{L}_i^{(m)}}
\begin{bmatrix}
f_{i,t}^{(\MP)}\\
f_{i,t}^{(\CI)}
\end{bmatrix}
+ \hdots +
\begin{bmatrix}
    \eta_{i,t}^{(I)}\\
    \eta_{i,t}^{(S)}
\end{bmatrix}
\quad\Leftrightarrow\quad
\begin{bmatrix}
    \epsilon_{i,t}^{(I)}\\
    \epsilon_{i,t}^{(S)}
\end{bmatrix} =
\bm{L}_i^{(m)}
\begin{bmatrix}
f_{i,t}^{(\MP)}\\
f_{i,t}^{(\CI)}
\end{bmatrix}
+
\begin{bmatrix}
    \eta_{i,t}^{(I)}\\
    \eta_{i,t}^{(S)}
\end{bmatrix}.
\end{equation*}
That is, we may use the factor loadings to identify the MP and CI shock by exploiting correlations in domestic instruments and the notion from basic economic and asset pricing theory that conventional monetary policy moves the interest rate and stock market (surprises) in opposite directions. By contrast, information shocks cause co-movement between the two. This is because the rate increase in the case of contractionary policy (associated with the central bank information shock) may reveal the confidence of the central bank regarding the economic outlook, thereby lifting the expectations of market participants. This idea for separating the two shocks goes back to \citet{jarocinski2020deconstructing}. The $+$ and $-$ signs above indicate these restrictions, and we formalize their implementation below.

In simple terms, the observed instruments $\bm{m}_{i,t}$ are a weighted sum of the structural shocks (encoded in the factors), plus a potentially predictable mean component and a measurement error. This specification is similar in spirit to the Bayesian proxy SVAR in \citet{caldara2019monetary} but differs along several key dimensions. We discuss these dimensions and additional details below. 

In this context, it is also again noteworthy that we impose these sign restrictions later on via a truncated normal prior on the loadings (see the matrix $\bm{L}_i^{(m)}$ in the equation above) instead of rotating the structural shocks. In conjunction with assuming heteroskedastic factors, this allows us to achieve point identification of the responses \citep[see][for details]{chan2022large}. This also means --- different from traditional approaches to sign restrictions \citep[see, e.g.,][]{arias2018inference}, which take an ``agnostic'' stance on structural parameters via exploiting the rotational invariance of the Gaussian distribution --- that we take an explicit stance about structural identification via informative priors, see also \citet{baumeister2015sign}.

\subsubsection*{Relationship to SVARs} 
Let $\bm{x}_t = (\bm{y}_{t-1}',\hdots,\bm{y}_{t-P}')'$, $\bm{A} = (\bm{A}_1,\hdots,\bm{A}_P)$, $\bm{B}_0$ is an invertible matrix of size $M$, and $\bm{A} = \bm{B}_0^{-1}\bm{B}$. The nexus between a conventional structural VAR (SVAR), such as $\bm{B}_0\bm{y}_t = \bm{B}\bm{x}_t + \bm{e}_t$ with $\bm{e}_t\sim\mathcal{N}(\bm{0},\bm{I}_M)$ referring to normalized structural shocks, and the reduced form given in Eq. (\ref{eq:pVAR}) and (\ref{eq:fsv}), is evident after some re-arranging: 
\begin{align*}
\bm{B}_0 \bm{y}_t &= &&\bm{B}\bm{x}_t + \bm{e}_t\\
{(\bm{L}'\bm{L})^{-1}\bm{L}'}\bm{y}_t &= {(\bm{L}'\bm{L})^{-1}\bm{L}'\bm{A}}\bm{x}_t + \bm{f}_t + (\bm{L}'\bm{L})^{-1}\bm{L}'\bm{\eta}_t \quad\approx\quad &&\tilde{\bm{B}}\bm{x}_t + \bm{f}_t,
\end{align*}
where $\tilde{\bm{B}} = (\bm{L}'\bm{L})^{-1}\bm{L}'\bm{A}$ and a central limit theorem implies that $(\bm{L}'\bm{L})^{-1}\bm{L}'\bm{\eta}_t \rightarrow \bm{0}$ asymptotically, see \citet{korobilis2022new} for additional discussions. This represents a reduced-rank SVAR model where $\bm{f}_t$ collects (uncorrelated) primitive structural shocks. The impulse response on impact in economy $i \in \{\EA,\US,\text{R}\}$ to shock $s\in\{\MP,\CI\}$ in economy $j \in \{\EA,\US\}$ thus is:
\begin{equation}
\frac{\partial \bm{y}_{i,t}}{\partial f_{j,t}^{(s)}} = \bm{l}_{i}^{(j,s,y)},\label{eq:impact}
\end{equation}
where $\bm{l}_{i}^{(j,s,y)}$ is the column of $\bm{L}_{i}^{(j,y)}$ associated with $f_{j,t}^{(s)}$. That is, domestic shock impacts and static spillovers are captured with identified part of the loadings matrix $\bm{L}$, which we uniquely identify using sign restrictions and heteroskedastic factors.

Two additional aspects are worth noting. First, our approach differs from standard implementations of proxy SVARs for two reasons: (i) we include the instruments directly in our model; (ii) we do not estimate the structural shocks by running IV-type auxiliary regressions on the reduced form VAR innovations. That is, we neither need invertibility (i.e., the reduced form errors being linear functions of the structural shocks), nor a specific VAR equation that is linked to the structural shock of interest. Second, our framework indeed is closer and can be related to what is known as an \textit{internal} instrument in IV approaches to estimation \citep[see, e.g.,][]{plagborg2021local,li2024local}. To see this, suppose $\tilde{N} = 1$ and there is only a single available instrument, so $S = Q_f = 1$. Assuming that this instrument is truly exogenous and fully unpredictable, we may set $L^{(m)} = 1$ and the associated measurement error $\eta_{1,t} = 0$. In this case, $m_t = f_t$, and the first column of $\bm{L}$ measures the contemporaneous relationships (in the conditional mean, see \citealp{huber2018bayesian} for an application of this approach) between the instrument and all remaining variables that are stacked in $\bm{y}_t$. This is equivalent (up to a proportionality constant due to the scaling of $m_t$) to including the instrument first in a standard VAR and using a Cholesky decomposition of the covariance matrix to identify the contemporaneous impacts. When there are multiple instruments for multiple shocks, e.g., $S = Q_f > 1$, a similar argument can be made, for uncorrelated instruments at least, when setting $\bm{L}^{(m)} = \bm{I}_S$.\footnote{The setup above allows us to identify distinct MP and CI shocks originating in the two economies, each instrument giving rise to a corresponding structural shock. In principle, however, one could also use this framework to extract an international factor, similar to the joint monetary/fiscal component as in \citet{huber2024general}. This is outside the scope of the present paper, and we leave this opportunity for future research.}

\subsubsection{Static and dynamic spillovers and spillbacks}
To illustrate the economic significance of the distinct blocks of loadings in $\bm{L}$, let us consider a CI shock in the US as a specific example. The domestic impact response of the endogenous macroeconomic and financial variables is given by the vector $\bm{l}_{\US}^{\US,\CI,y}$, which we have identified by imposing joint zero and sign restrictions on the first $Q_f$ rows of $\bm{L}$. Note that these restrictions are exclusively related to the instruments. Since $\bm{l}_{\EA}^{\US,\CI,y}$ and $\bm{l}_{\text{R}}^{\US,\CI,y}$ measure the contemporaneous impact of a CI shock originating in the US on non-US variables (i.e., those of the EA and the rest of the economies considered, respectively), they are static spillovers of a foreign shock from the perspective of all these non-US economies.

Equation (\ref{eq:impact}) defines the response on impact for $h = 0$. Projecting the impact response forward $h$ steps via the dynamic coefficients in $\bm{A}$ yields the sequence of dynamic causal effects $\partial \bm{y}_{i,t+h}/\partial f_{j,t}^{(s)}$ across horizons $h = 1,2,\hdots,$ which in our PVAR setup then naturally also encompass the dynamic spillovers (and spillbacks) due to the iterative nature of the model. 

To see this more clearly, it is worthwhile to discuss which blocks of coefficients govern which types of dynamics explicitly. From the perspective of the US, the matrices $\bm{A}_{p[\US,\US]}^{(yy)}$ govern domestic shock transmission, i.e., how the impact measured by $\bm{l}_{\US}^{\US,\CI,y}$ propagates through domestic variables. By contrast, the matrices $\bm{A}_{p[\EA,\US]}^{(yy)}$ measure the sensitivity of EA variables to lagged dynamics in US variables, i.e., dynamic spillovers from US to EA variables. In addition, if a CI shock originating in the US yields a non-zero response at some horizon in the EA (or some other economy), the relationship between domestic US variables and the non-domestic EA lagged dynamics is determined by $\bm{A}_{p[\US,\EA]}^{(yy)}$. In this sense, a movement caused by a US shock in a non-domestic variable may ``spillback'' via the foreign quantities into its country of origin. These latter (non-domestic) and higher-order transmission channels are excluded implicitly in, e.g., the framework of \citet{jarocinski2022central}.

To give an economic example of why these different types of spillovers may exist, consider the following. A CI shock in the US affects both domestic and European market participants' expectations of a positive development of the US economy (which materializes, among other things, as a rise in stock markets due to higher expected returns). Due to international linkages between firms, market participants may expect that European suppliers of US firms will also benefit from this positive outlook on the US, which would be reflected in a simultaneous increase in European stock prices (these effects would be captured as static spillovers). This aggregate increase in stock prices may improve firms' financing conditions and, with a time lag, increase for example industrial production in the US (domestic transmission), and industrial production in the EA through ``downstream'' supply chain effects (dynamic spillovers); positive developments in the EA may then in turn affect price dynamics in the US (e.g., via increased EA demand for US goods, i.e., dynamic spillbacks), and so on.

\subsection{Priors, posteriors and sampling algorithm}
Our prior setup is inspired by the recent literature on high-dimensional (P)VARs. Most of our choices are standard, and we discuss them below. The first crucial and perhaps least well-known ingredient --- needed to achieve identification of the structural shocks --- is the prior used to impose the sign restrictions on some of the non-zero elements of the matrix of factor loadings.

Let $\bm{l}_i$ denote the $i$th row and $l_{ij}$ the $(i,j)$th element of the matrix $\bm{L}$ with $i = 1, \hdots, M,$ and $j = 1, \hdots, Q$. Each element of this matrix has an associated pre-defined variable $r_{ij} \in \{1, -1, 0, \text{NA}\}$, which we store in the $M\times Q$-matrix $\bm{R}$. This matrix signals the respective restrictions, positive ($+$), negative ($-$), zero, or none. Details about the positions of these restrictions in the matrix $\bm{L}$ were provided in the preceding sub-section. We may impose these explicitly with a prior of the form:
\begin{equation*}
    {l}_{ij}\sim
    \begin{cases}
	    \mathcal{N}(0, 1)\cdot\mathbb{I}({l}_{ij} \geq 0), &\text{if } r_{ij} = 1,\\
	    \mathcal{N}(0, 1)\cdot\mathbb{I}(l_{ij}\leq0), &\text{if } r_{ij} = -1,\\
	    0, &\text{if } r_{ij} = 0,\\
	    \mathcal{N}(0, 1),&\text{if } r_{ij} = \text{NA}.
    \end{cases},
\end{equation*}
where we use (truncated) normal distributions with variance equal to $1$. From a sampling perspective, we may update each row of the matrix $\bm{L}$ independently. This prior setup yields posteriors that are either truncated normal or normal distributions from which samples can be obtained efficiently; the zero restrictions can trivially be imposed by excluding the respective factor when sampling the remaining non-zero loadings for each endogenous variable. For a detailed theoretical and econometric discussion of how this achieves identification in conjunction with heteroskedastic factors, see \citet{chan2022large}.

For the variances of the latent factors collected on the diagonal of $\bm{H}_t$, we use a standard SV specification, with independent AR(1) state equations:
\begin{equation*}
    h_{j,t} = \phi_{h,j} h_{j,t-1} + \varsigma_{h,j} e_{j,t}, \quad e_{j,t} \sim \mathcal{N}(0,1), \quad \text{for } j = 1,\hdots,Q.
\end{equation*}
Note that the unconditional means of these processes are normalized to zero to pin down the scaling of the latent factors. Turning to the parameters governing these state equations, we rely on a Beta-distributed prior on the $\phi_{h,j}$'s which favors persistent processes, and on the $\varsigma_{h,j}$'s we use mildly informative Gamma priors. We assume that the idiosyncratic measurement errors are homoskedastic, and use independent inverse Gamma priors $\omega_{i}^2 \sim \mathcal{G}^{-1}(a_{\omega,i},b_{\omega,i})$ for $i = 1,\hdots, M$. On the errors associated with the instruments, we impose the belief that measurement errors are usually small (specifically, $a_{\omega,i} = 30$ and $b_{\omega,i} = 0.3$); on the remaining equations, the prior is set to be less informative (with $a_{\omega,i} = 3$ and $b_{\omega,i} = 0.3$).

On the VAR coefficients, we use a variant of a global-local shrinkage prior, adapted to the panel structure of our dataset. Let $\bm{\alpha}_{\text{dom}}$ denote a vector which stacks all parameters in $\bm{A}$ that govern domestic dynamics, i.e., $\bm{A}_{p[ii]}^{(y\bullet)}$; and $\bm{\alpha}_{\text{for}}$ is a vector that stores those associated with non-domestic (foreign) variables, i.e., those in the matrices $\bm{A}_{p[ij]}^{(y\bullet)}$ where $i \neq j$. We use the conditionally independent hierarchical Gaussian priors:
\begin{equation*}
    \alpha_{d,k} \sim \mathcal{N}(0, \tau_{d}^2\lambda_{d,k}^2), \quad \text{for } d \in \{\text{dom},\text{for}\}, \quad k = 1,\hdots,
\end{equation*}
where $k$ indexes the elements in the respective stacked vectors. Here, $\tau_{d}^2$ is a global shrinkage factor that governs the overall degree of shrinkage, whereas $\lambda_{d,k}^2$ are local scaling parameters that may detect important variables even in the presence of strong global shrinkage (which typically arises in high-dimensional systems). 

Instead of relying on a truly ``{global}'' shrinkage parameter (pooling across domestic and foreign dynamics), having distinct $\tau_{\text{dom}}^2$ and $\tau_{\text{for}}^2$ allows for different amounts of global shrinkage on distinct subsets of parameters. This reflects the notion that domestic dynamics are usually more important than non-domestic ones \citep[see][for a discussion and empirical evidence for this claim]{feldkircher2022approximate}. So, typically, we would expect $\tau_{\text{for}}^2 \ll \tau_{\text{dom}}^2$ and allow for this case, but we do not need to explicitly impose it. In this context, it is also worth noting that excluding non-domestic variables deterministically may lead to omitted variable bias in a given domestic economy. Our prior is in fact designed to detect important domestic and non-domestic variables automatically and shrink noise in a data-driven fashion. Specifically, among the class of global-local priors, we choose the horseshoe (HS) prior, see \citet{carvalho2010horseshoe}, due to its good empirical properties and lack of tuning parameters. This prior arises when choosing $\tau_{d},\lambda_{d,k}\sim\mathcal{C}^{+}(0,1)$, for $k = 1,\hdots,$ where $\mathcal{C}^{+}$ is the half-Cauchy distribution.\footnote{Note that to avoid affecting shrinkage patterns by different scalings of the underlying data, we standardize all variables before estimation to have zero mean and unit variance as is common in a penalized regression or shrinkage context, and transform all results of interest back to the original scale ex-post.}

On the elements of the matrices $\bm{A}_{p[ij]}^{(m\bullet)}$, i.e., those associated with the dynamic behavior of the instruments, we use informative independent zero-mean Gaussian priors with a variance $10^{-6}$ reflecting the belief that the instruments are unpredictable (we note that data information may still overrule this prior and unpredictability is not imposed dogmatically). By contrast, we use vague Gaussian priors with variance $10$ on all deterministic terms featured in the conditional mean of our model.

\subsubsection*{Sampling Algorithm}
From a sampling perspective, we note that our assumptions, especially those in the context of the factor model, allow us to update virtually all quantities equation-by-equation, factor-by-factor, or variable-by-variable, which offers significant computational advantages. Below we provide an overview of our sampling algorithm:
\begin{enumerate}[leftmargin = *]
    \item The VAR coefficients can be updated equation-by-equation conditional on the factors and loadings. Let $\tilde{y}_{i,t}$ denote the $i$th element of the $M$-vector $\tilde{\bm{y}}_t = \bm{y}_t - \bm{L}\bm{f}_t = \bm{A}\bm{x}_t + \bm{\eta}_t$, then the $i$th VAR equation is given by:
    \begin{equation*}
        \tilde{y}_{i,t} = \bm{A}_{i,\bullet}\bm{x}_t + \eta_{i,t}, \quad \eta_{i,t} \sim\mathcal{N}(0, \omega_{i}^2),
    \end{equation*}
    where $\bm{A}_{i,\bullet}$ is the $i$th row of $\bm{A}$. That is, we have $M$ independent homoskedastic linear regressions due to $\bm{\Omega}$ being a diagonal matrix. The posterior of $\bm{A}_{i,\bullet}$ is multivariate Gaussian with its prior covariances defined by the HS prior, of $\omega_{i}^2$ it is inverse Gamma, and the corresponding posterior moments can be found in any Bayesian textbook that discusses linear regression.

    \item Looping through equations yields a full set of draws for $\bm{A}$. Subsetting these according to the domestic and foreign vectors $\bm{\alpha}_{\text{dom}}$ and $\bm{\alpha}_{\text{for}}$ allows for straightforward updates of the global and local shrinkage factors, see \citet{makalic2015simple}.
    
    \item The loadings can also be sampled equation-by-equation, by obtaining the $i$th reduced form error conditional on the VAR coefficients as $\epsilon_{i,t} = \bm{l}_i\bm{\mathrm{f}}_t + \eta_{i,t}$. Stacking over $t = 1, \hdots, T$, we have $\bm{\epsilon}_i = \bm{\mathrm{F}}\bm{l}_i + \bm{\eta}_{i}$. The posterior of the loadings is:
    \begin{align*}
        \bm{l}_i|\bm{r}_i,\bullet &\sim \mathcal{N}({\bm{\mu}_{\bm{l}_i}},{\bm{V}_{\bm{l}_i}})\cdot\mathbb{I}(\underline{\bm{l}_i} < \bm{l}_{i} < \overline{\bm{l}_i}),\\
        {\bm{V}_{\bm{l}_i}} &= (\omega_i^{-2}\bm{\mathrm{F}}'\bm{\mathrm{F}} + \bm{I}_Q)^{-1}, \quad {\bm{\mu}_{\bm{l}_i}} = {\bm{V}_{\bm{l}_i}}(\omega_i^{-2}\bm{\mathrm{F}}'\bm{\epsilon}_i),
    \end{align*}
    which is a truncated normal distribution, truncated to the support implied by the restrictions in the $i$th row of $\bm{R}$, $\bm{r}_i$, which yields the respective lower ($\underline{\bm{l}_i}$) and upper ($\overline{\bm{l}_i}$) bounds.\footnote{The elements of these bounds, $(\underline{l_{ij}},\overline{l_{ij}})$, are set as follows: if $r_{ij} = 1\rightarrow(0,\infty)$, if $r_{ij} = -1\rightarrow(-\infty,0)$, if $r_{ij} = \text{NA}\rightarrow(-\infty,\infty)$; if $r_{ij} = 0\rightarrow(0,0)$ which indicates deleting the corresponding dimension.} In case there are no restrictions, the posterior is a normal distribution.

    \item The factors $\tilde{\bm{\mathrm{F}}} = (\bm{\mathrm{f}}_1',\hdots,\bm{\mathrm{f}}_T')'$ can be sampled jointly by defining $\tilde{\bm{L}} = (\bm{I}_T \otimes \bm{L})$, $\tilde{\bm{\epsilon}} = (\bm{\epsilon}_1,\hdots,\bm{\epsilon}_T)'$, $\tilde{\bm{\Omega}} = (\bm{I}_T \otimes \bm{\Omega})$ and $\tilde{\bm{H}} = \text{bdiag}(\bm{H}_1,\hdots,\bm{H}_T)$. In this case, Eq. (\ref{eq:fsv}) can be written in seemingly unrelated regression form, and textbook results for linear regressions can again be used to obtain the posterior $\tilde{\bm{\mathrm{F}}}|\bullet\sim\mathcal{N}(\bm{\mu}_{\tilde{\bm{\mathrm{F}}}}, \bm{V}_{\tilde{\bm{\mathrm{F}}}})$ with moments:
    \begin{equation*}
        \bm{V}_{\tilde{\bm{\mathrm{F}}}} = (\tilde{\bm{L}}'\tilde{\bm{H}}^{-1}\tilde{\bm{L}} + \tilde{\bm{\Omega}}^{-1})^{-1},\quad \bm{\mu}_{\tilde{\bm{\mathrm{F}}}} = \bm{V}_{\tilde{\bm{\mathrm{F}}}}(\tilde{\bm{L}}'\tilde{\bm{H}}^{-1}\tilde{\bm{\epsilon}}).
    \end{equation*}

    The obtained factors $\{\mathrm{f}_{q,t}\}_{t=1}^{T}$ for $q = 1,\hdots,Q,$ can subsequently be used to update the volatility processes and state equation parameters using a standard SV algorithm factor-by-factor, see \citet{kim1998stochastic}.
\end{enumerate}

\section{Data and model specification}\label{sec:data}
Our model is estimated based on a panel of monthly data ranging from 1999:01 to 2019:12. We use $P = 12$ lags and $Q_{\tilde{f}} = 8$ additional factor alongside the instruments.\footnote{Our dataset ends before the COVID-19 pandemic to avoid having to introduce additional model features to control for the corresponding outliers during this period.}

The instruments for structural shocks in Eq. (\ref{eq:pVAR}) are constructed in line with \citet{jarocinski2020deconstructing}. Specifically, we use the high-frequency surprises in interest rates and stock markets for the EA and the US. The data are provided on the \href{https://marekjarocinski.github.io}{webpage of Marek Jaroci\'nski}. The interest rate surprises for the EA are the first principal component of the monetary event-window changes in overnight index swaps (OIS) with maturities one-, three-, six-months and one-year (identifiers OIS1M, OIS3M, OIS6M, OIS1Y, see also \citealp{altavilla2019measuring}); the stock market surprises are changes in the Euro Stoxx 50 during the monetary event-window. For the US, the interest rate surprises are the first principal component of the surprises in interest rate derivatives with maturities from one month to one year (identifiers MP1, FF4, ED2, ED3, ED4, see also \citealp{gurkaynak2005actions}); and stock market surprises are based on changes in the S\&P500 in a 30-minute window surrounding Federal Open Market Committee (FOMC) announcement dates.

For the set of US and EA aggregate quantities, we rely on various data sources, provided by the OECD, the FRED database, the ECB SDW and Macrobond. As key policy rates, we follow \citet{jarocinski2020deconstructing} and use the one-year US treasury yield, and the one-year Bund yield as the safest EA interest rate. To capture stock market reactions, we include the S\&P 500 and the Euro Stoxx 50 index. As a monthly measure of economic activity, we rely on year-on-year industrial production growth, while inflation is measured by year-on-year growth rates of headline consumer price indexes. For measuring financial conditions, we rely on the excess bond premium \citep[EBP,][]{gilchrist2012credit} for the case of the US, and the option-adjusted spread (OAS) of the ICE BofA Euro High Yield Index in the EA. The set of variables for the aggregate equations is completed by the USD/EUR exchange rate and oil prices to capture international factors. A summary of the employed series, transformations, and sources of the data is provided in the Appendix.

The set of country-specific macroeconomic and financial variables in $\bm{\mathrm{y}}_{\text{R},t}$ comprises the following series downloaded from the FRED database and the OECD. We include inflation in consumer prices, while industrial production growth serves as a monthly indicator of economic activity. The data set includes observations for Canada (CA), Japan (JP), and the United Kingdom (UK). 

\section{Empirical results}\label{sec:results}
\subsection{Revisiting transatlantic spillovers}
\subsubsection{Identified structural factors}
\begin{figure}[t]
    \includegraphics[width = \textwidth]{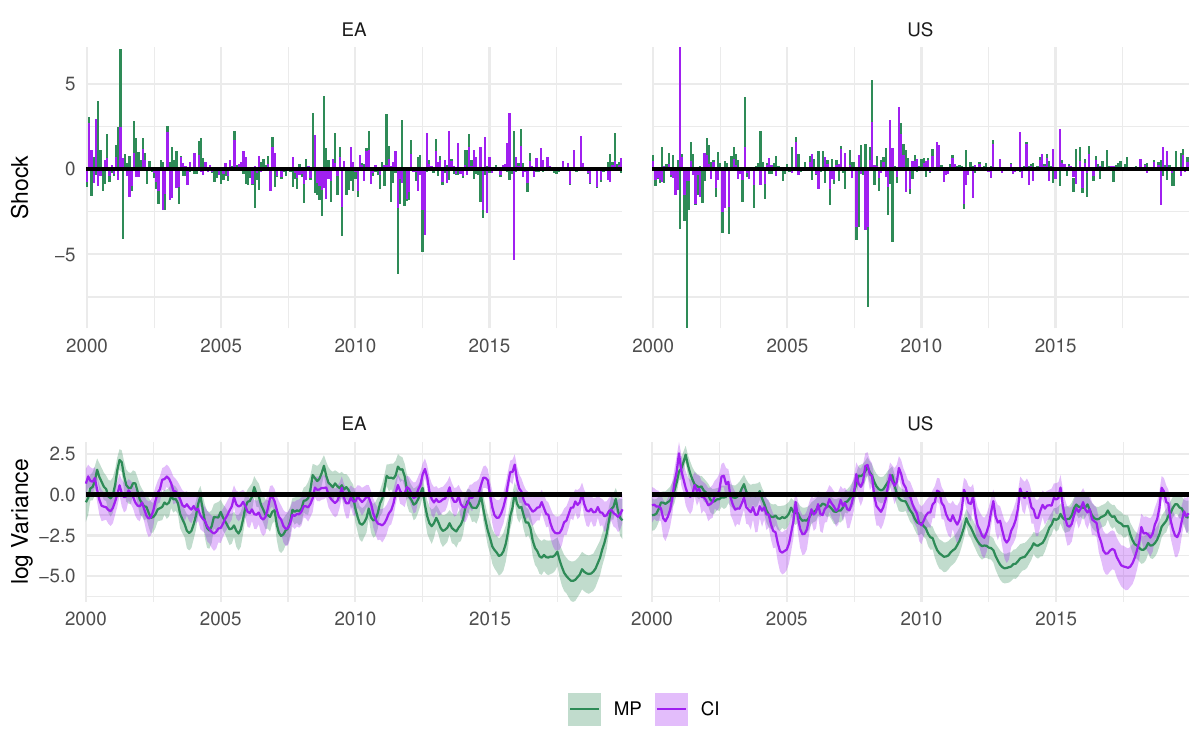}
    \caption{Posterior median estimate of the identified monetary policy (MP) and central bank information (CI) shocks for the EA and US (upper panel) and associated log volatility of the respective shocks (lower panel).}\label{fig:shocks}
\end{figure}

We begin the discussion of our empirical results by exploring the identified structural MP and CI shocks in the EA and the US. The upper panels of Figure \ref{fig:shocks} show the posterior median estimates of the identified factors $\bm{f}_t$. The lower panels show the posterior median and 68 percent posterior credible sets of the associated log volatility processes. The charts in the upper panels are inspired by those shown in \citet[][Figs. 4 and 9]{jarocinski2020deconstructing}. There are some differences compared to the original paper, but by and large, our approach to identifying MP and CI shocks in a multicountry context seems to yield comparatively similar shock series, even though our approach to identification is rather different. Specifically, deviations can be caused by several factors, the shocks being extracted from larger information sets that span multiple economies, and the presence of measurement errors among them.

The lower panels of Figure \ref{fig:shocks}, by contrast, provide novel insights. Recall that to identify the scale of the latent factors, we normalize their unconditional variance to $1$, which yields an unconditional mean of $0$ for the log volatilities. In the context of our impulse response functions below, this also means that our estimates reflect a one standard deviation shock in the respective factors (which by construction is equivalent to a unit shock). The volatilities shed light on several aspects of monetary shocks over time. Specifically, there are distinct patterns that hint at the varying importance of the respective shock. For instance, when approaching the effective lower bound (ELB), intuitively, the variance of the monetary policy shock is smaller than in other periods. In contrast, the variance of information shocks remains more stable during those periods, pointing towards their increased relative importance when monetary policy was faced with the ELB. This discussion relates to \citet{amir2022understanding}, who find that often used instruments, such as ours, usually are informative only during specific economic episodes, and identification of any causal effects is driven chiefly by these periods.

\subsubsection{Investigating static impact spillovers}
Before turning to the dynamic effects of the identified shocks, it is worthwhile to assess the responses of the endogenous variables on impact. As discussed in detail in Section \ref{subsec:identifyingstrufac}, our approach to identification boils down to giving economic meaning to the factors we identify with a mix of external information, sign/zero restrictions, and heteroskedasticity. The corresponding factor loadings then measure the impact of the respective structural shock at $h = 0$. Table \ref{tab:impact} displays these estimates, for the EA and US variables, in the form of the posterior median alongside the $68$ percent posterior credible set in parentheses. Blue and red shading indicates significant positive and negative estimates (defined as the $68$ percent credible set not including $0$), respectively. Insignificant impacts are shaded in gray. 

\begin{table}[t]
    \caption{Impact effects at $h = 0$ of the indicated shocks as measured by the blocks of loadings $\bm{L}_{i}^{(m)}$. Posterior median and 68 percent posterior credible set in parentheses.}\label{tab:impact}
    \includegraphics[width = \textwidth]{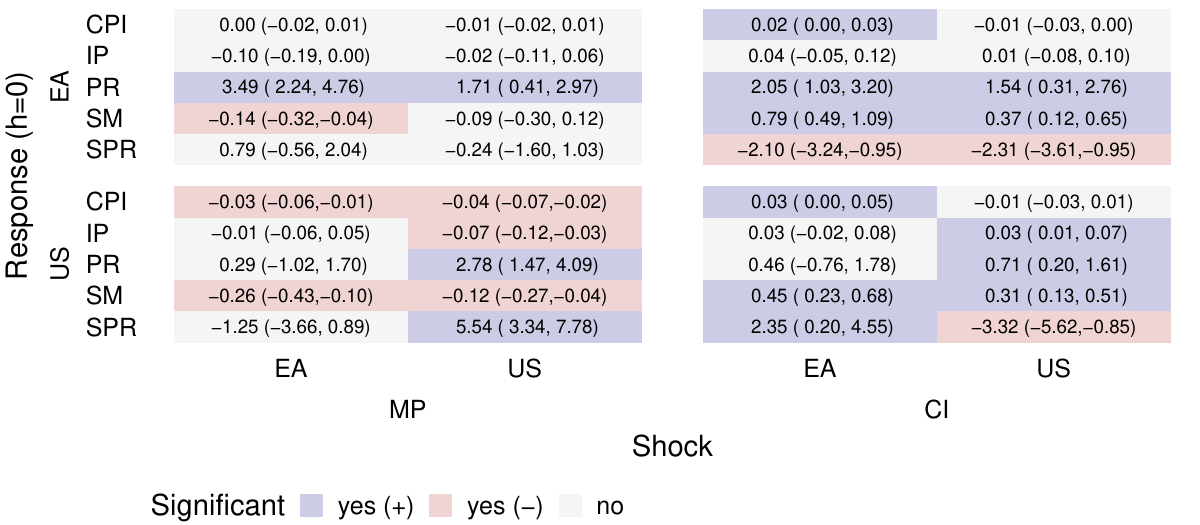}
\end{table}

It is worth reiterating that these estimates are the identified contemporaneous responses. That is, we model both domestic shock impacts (e.g., the response of the EA policy rate to a EA monetary policy shock), but also, static spillovers (e.g., the response of the US policy rate to an EA monetary policy shock). Table \ref{tab:impact} is structured such that the rows indicate these impacts for each one of the EA and US variables, while the columns refer to the shock origin (ECB for EA; and Fed for US) and type (monetary policy, MP; and central bank information, CI). 
The variables include inflation in the consumer price index (CPI), industrial production growth (IP), the policy rate (PR), stock market index (SM), represented by the S\&P 500 for the US and Euro Stoxx 50 for the EA, and spreads (SPR), which reflect broader financial conditions.

Starting with domestic responses, we briefly characterize the magnitude of the impact effects. The domestic monetary policy shocks increase the respective domestic policy rate (PR) by about $3.5$ basis points (BPs) in the EA and $2.8$ BPs in the US. These comparatively small impacts are unsurprising, given that a significant amount of our sampling period was affected by the zero/effective lower bound. For the central bank information shock, we have an increase of the PR in the EA at about $2$ BPs, while in the US it increases by about $1$ BP in response to the domestic shock.\footnote{We note that these impacts could be rescaled to reflect different shock sizes and impacts on specific endogenous variables (the underlying VAR is linear and responses are symmetric, so the scaling is a mere normalization) --- we rely on the one-standard deviation interpretation in terms of the latent shocks to help comparability across shock types and origins.} The domestic estimates agree by and large with the related literature, e.g., our estimates qualitatively and proportionally agree with those of \citet{jarocinski2020deconstructing}. Our alternative approach to identification thus yields the expected signs of the responses for the respective domestic policy rate and stock market.

Consistent also with \citet{jarocinski2022central}, our findings suggest that CI shocks originating in the EA and the US cause significant spillovers more frequently than MP shocks --- visually, this can be inspected by noticing that there are many more colored cells in the right panels of Table \ref{tab:impact} than the left panels. For IP, most static spillover estimates are either statistically insignificant or small in magnitude, while for CPI, there are some modestly sized contemporaneous effects of non-domestic shocks. Thus, a clear and important pattern that emerges is that contemporaneous spillovers are stronger and more frequent for financial rather than macroeconomic variables. Shocks of both types and origins typically affect, significantly and meaningfully, the corresponding non-domestic policy rates, stock markets, and spreads, corroborating the findings of \citet{miranda2022tale}. It is worth noting that for the CI shocks, spreads (SPR) seem to be the quantitatively more important contemporaneous transmission channel, rather than the policy rate.

To sum up this discussion of static spillovers, we find that our estimates corroborate those found in the preceding literature. Notably, \citet{jarocinski2022central} undertakes a similar study to ours and includes the shocks originating in the EA in a VAR featuring solely US variables (and vice versa). This approach is related to ours but abstracts from potential higher-order spillover (and spillback) effects that arise from the dynamic evolution of a fully specified multicountry VAR, see our discussion in Section \ref{subsec:identifyingstrufac}. For instance, the US monetary policy shock might decrease US economic activity and income, thereby (besides domestic demand) decreasing US demand for EA goods. This in turn would affect the EA economic outlook and inflation (already reflected as expectations in financial markets, more or less contemporaneously), but with a pronounced time lag in real variables and prices. Our PVAR allows us to capture these higher-order responses explicitly, and we will turn to estimates of these dynamic effects next.

\subsubsection{Dynamic higher-order effects and spillovers}
\begin{figure}[t]
    \begin{subfigure}[t]{0.49\textwidth}
    \caption{Euro area (EA)}
    \includegraphics[width=\textwidth]{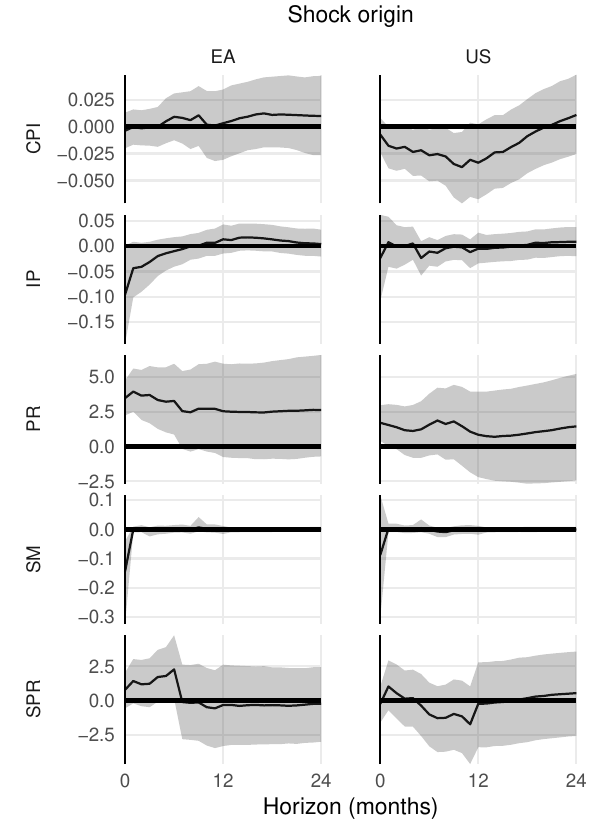}
    \end{subfigure}
    \begin{subfigure}[t]{0.49\textwidth}
    \caption{United States (US)}
    \includegraphics[width=\textwidth]{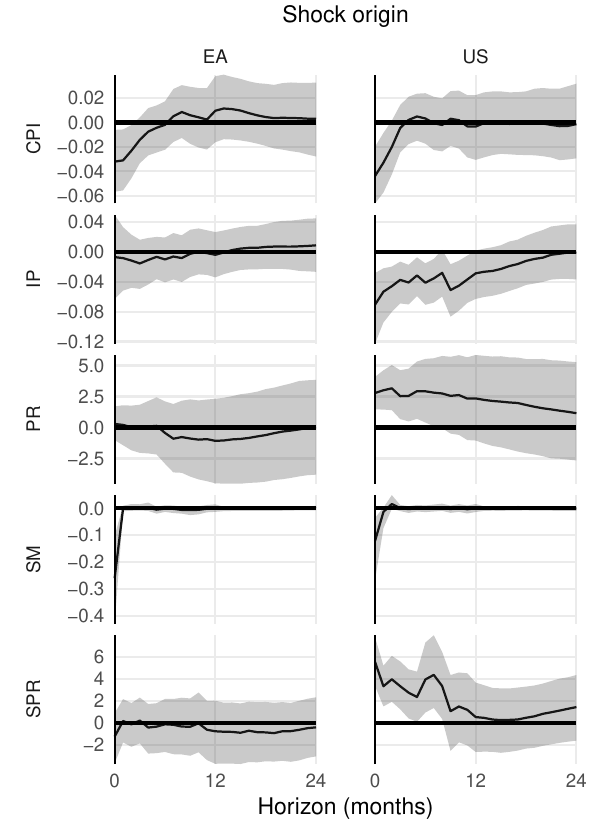}
    \end{subfigure}
    \caption{Responses of the indicated variables to a monetary policy shock. Posterior median alongside the 68 percent credible set.}
    \label{fig:mp-shock_domfor}
\end{figure}

The following set of results presents dynamic higher-order effects and cross-country spillovers of monetary policy and central bank information shocks. We note that our econometric framework features an explicit model of possible dynamic spillovers and spillbacks, which distinguishes our findings from those of \citet{jarocinski2022central}. The first set of dynamic impulse responses in Figure \ref{fig:mp-shock_domfor} shows the effects of a contractionary monetary policy shock in the EA and the US as well as the respective spillovers. Panel (a) displays responses of EA variables to a monetary policy shock originating from the EA (left column) and the US (right column), panel (b) follows the same pattern for responses in US variables. The charts depict the posterior median with the 68 percent credible set in impulse response functions.

\begin{figure}[t]
    \begin{subfigure}[t]{0.49\textwidth}
    \caption{Euro area (EA)}
    \includegraphics[width=\textwidth]{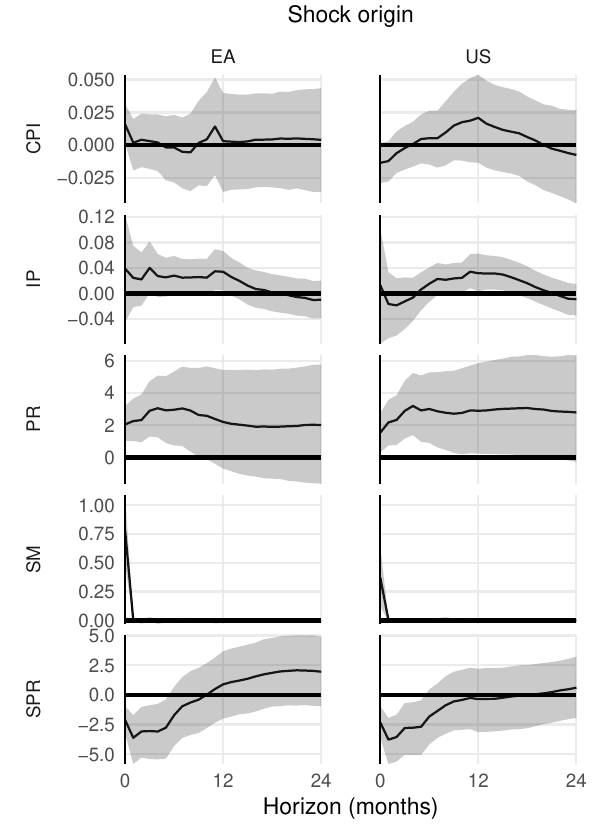}
    \end{subfigure}
    \begin{subfigure}[t]{0.49\textwidth}
    \caption{United States (US)}
    \includegraphics[width=\textwidth]{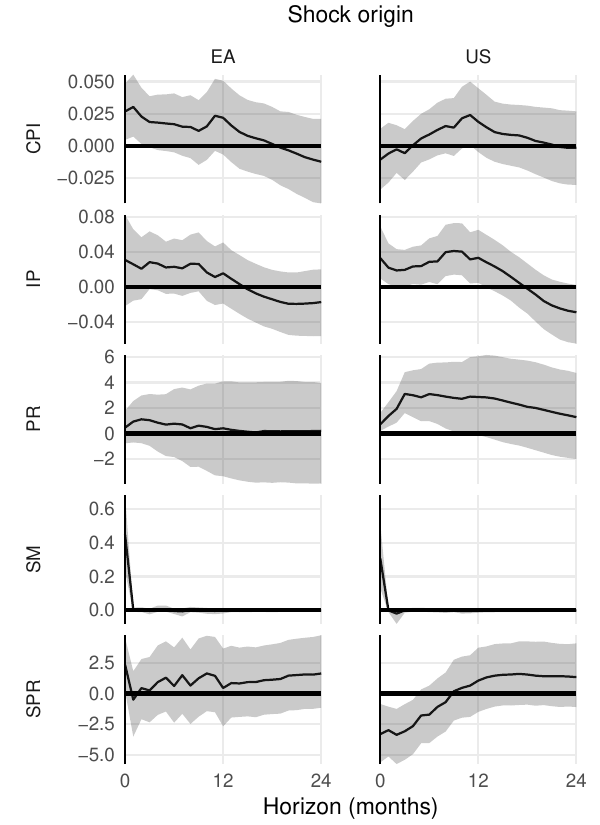}
    \end{subfigure}
    \caption{Responses of the indicated variables to a central bank information shock. Posterior median alongside the 68 percent credible set.}
    \label{fig:ci-shock_domfor}
\end{figure}

The figure shows that a monetary policy shock in each economy has varying effects on these indicators. A contractionary monetary policy shock originating in the EA has no significant effect on CPI inflation in the EA, which remains flat virtually over all horizons. This finding of limited effectiveness of changes in the policy rate might be connected with the decreasing variance (in our sampling period) of monetary policy shocks previously mentioned and could be explained by the decreasing effectiveness of conventional monetary policy when approaching the ELB \citep[see, e.g.,][]{borio2017monetary}. A further explanation for this finding are state-dependent asymmetries in monetary policy effects \citep{tenreyro2016}, or masked heterogeneity in effects between different EA member countries \citep{ciccarelli2013heterogeneous, de2013bank, doi:10.1111/obes.12038, georgiadis2015examining}. In contrast, we find that a tightening in EA monetary policy has a short-lived disinflationary effect on the US.\footnote{Figures \ref{fig:mp-shock_domfor_zoom} and \ref{fig:ci-shock_domfor_zoom} provide a closer look at the impulse responses for the first six months after an MP or CI shock occurred, respectively.}

By contrast, CPI inflation responses to a monetary tightening in the US are more in line with theoretical predictions. While decreases in inflation are relatively short-lived in the US, spillovers to EA inflation show a more persistent effect, although the response is significantly different from zero only for a limited period. Industrial production growth shows a more pronounced decline in the EA, reflecting reduced economic output following a policy tightening. While the response of EA IP is not statistically significant, growth in US IP declines significantly and persistently for almost a year after the monetary policy shock occurred. 

Turning to financial variables, the policy rate rises in response to the domestic monetary policy shocks, the effect being more persistent in the EA. Our results also show that domestic monetary policy has a stronger impact on interest rates than spillover effects. It is worth pointing out that we see a more pronounced reaction in the EA policy rate to a US monetary policy tightening than the over way around, which points towards the global importance of US economic shocks. Domestic stock market returns experience a short-lived decline in both economies, highlighting immediate financial market reactions to policy adjustments. While similarly short-lived, we also find monetary policy spillovers to financial markets on both sides of the Atlantic, reflecting the global and fast-paced character of the financial system. The relationship between monetary policy in the US and its consequences for global equity markets is well researched \citep{ehrmann2009global, hausman2011global, cesa2022financial}. Our findings indicate that EA monetary policy might play a similarly important role. Lastly, spreads widen in response to the shock, indicating tightening financial conditions. In contrast to the financial market indicators, spreads seem to be driven more by domestic factors, as we can see little to no spillover effects on these variables.

Summing up, real and financial variables respond to a domestic monetary policy tightening mostly as expected and our findings are roughly in line with results by \citet{jarocinski2020deconstructing}, who consider the US and EA individually. Spillovers from monetary policy tend to be modest, with US shocks having larger effects on the EA than vice versa, a result that is roughly in line with findings by \citet{ca2023making}. One explanation for this is that we see a stronger adjustment in the policy rate of the ECB in response to a tightening in the US. This (conscious or unconscious) coordination/synchronization results in a tightening and relatively stronger higher-order effects (than vice versa) in the EA as well.

Figure \ref{fig:ci-shock_domfor} depicts the responses of EA and US variables to domestic and foreign central bank information shocks. In the EA, domestic shocks elicit only very limited movement in CPI inflation. US inflation responds more strongly to a domestic CI shock, and interestingly we find that CI shocks originating in the EA also significantly spill over to US prices for a short time after impact. In contrast, inflation in the EA displays an initial and barely significant decrease after a US CI shock, followed by a rebound and (insignificant) peak around one year after impact. IP growth in the EA reacts very similarly to a CI shock regardless of whether the shock originates in the EA or the US, with a persistent increase until about one year after the impact. The peak occurs after about a year and is statistically significant in both cases. Turning to the US, IP growth responds positively to a domestic shock and the hump-shaped response function is significant for more than a year; spillovers from the EA tend to provoke a positive response also in US CPI inflation, but this response is not statistically significant over the horizons we consider.

In terms of financial transmission channels, the policy rate also shows positive and persistent responses to domestic CI shocks, for both the EA and the US. While the US policy rate shows no reaction to EA CI shocks, policy rates in the EA react positively to a US shock, quite similarly in magnitude to the domestic shock. Stock markets display a positive response to CI shocks, no matter the origin, which is, however, similarly short-lived than the reaction to a monetary policy shock. This result is another indication of the way financial markets react very swiftly to news from both the EA and the US. A domestic CI shock further lowers spreads in the EA and the US, in line with an easing in financial conditions and an improved economic outlook. We also observe that EA spreads narrow in response to a US CI shock, while similar spillovers from the EA to the US are absent.

\begin{figure}[t]
    \begin{subfigure}[t]{0.49\textwidth}
    \caption{Monetary policy shock}
    \includegraphics[width=\textwidth]{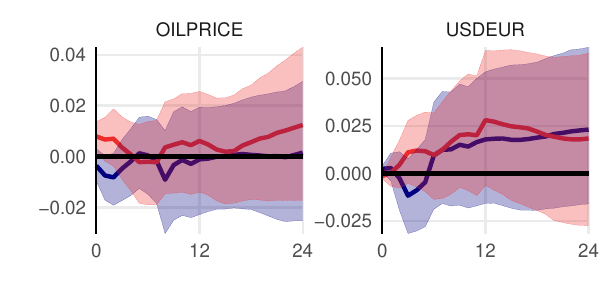}
    \end{subfigure}
    \begin{subfigure}[t]{0.49\textwidth}
    \caption{Central bank information shock}
    \includegraphics[width=\textwidth]{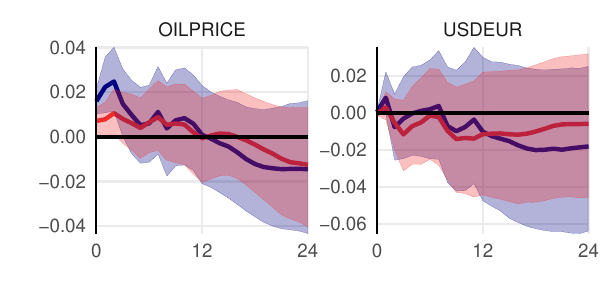}
    \end{subfigure} 
    \caption{Responses for international variables to the respective shock in the EA (blue) and the US (red). Posterior median alongside the 68 percent credible set.}
    \label{fig:ci-shock_inter}
\end{figure}

Figure \ref{fig:ci-shock_inter} extends the analysis by showing the impact of monetary policy and central bank information shocks on ``international variables,'' namely oil prices (OILPRICE) and the USD/EUR exchange rate (USDEUR). The left panel focuses on a standard monetary policy shock, while the right examines central bank information shocks. Both US shocks have an upward effect on oil prices which dies out a few months after impact, but have no significant effect on the exchange rate. In contrast, EA monetary policy shocks have a smaller and insignificant effect on both oil prices and the exchange rate. Central bank information shocks in the EA appear to have comparable effects on oil prices and currency movements to those originating from the US, resulting in a short-lived increase in oil prices and no significant response in the exchange rate after a CI shock. It is interesting to note that, again, information shocks from the EA tend to influence the global economy more strongly than pure monetary policy shocks, while for the US, both monetary policy and CI shocks result in responses of similar magnitude. These results underscore the asymmetric influence of US and EA central bank actions on global markets.

\subsection{Spillovers to other economies}
\begin{figure}[t]
    \begin{subfigure}[t]{0.49\textwidth}
    \caption{Monetary policy shock}
    \includegraphics[width=\textwidth]{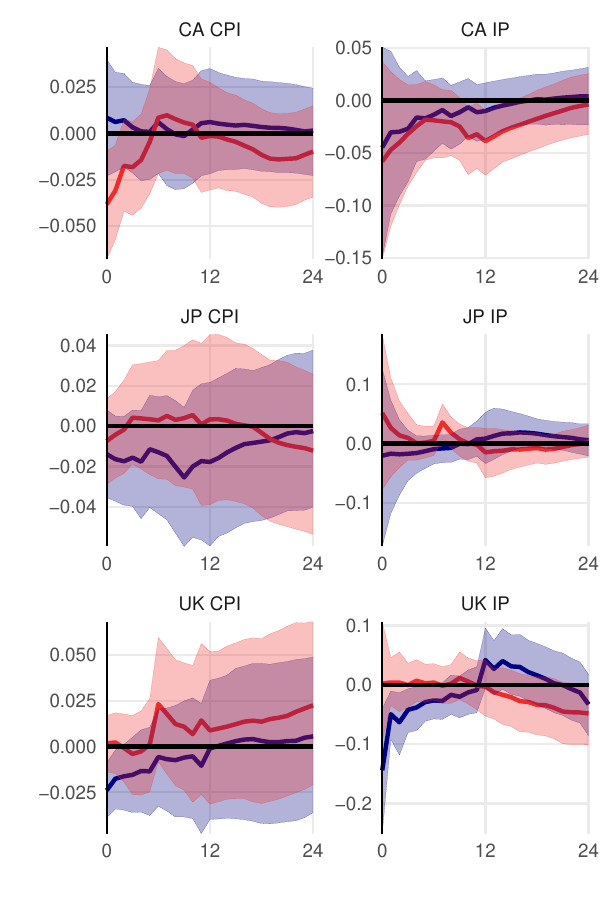}
    \end{subfigure}
    \begin{subfigure}[t]{0.49\textwidth}
    \caption{Central bank information shock}
    \includegraphics[width=\textwidth]{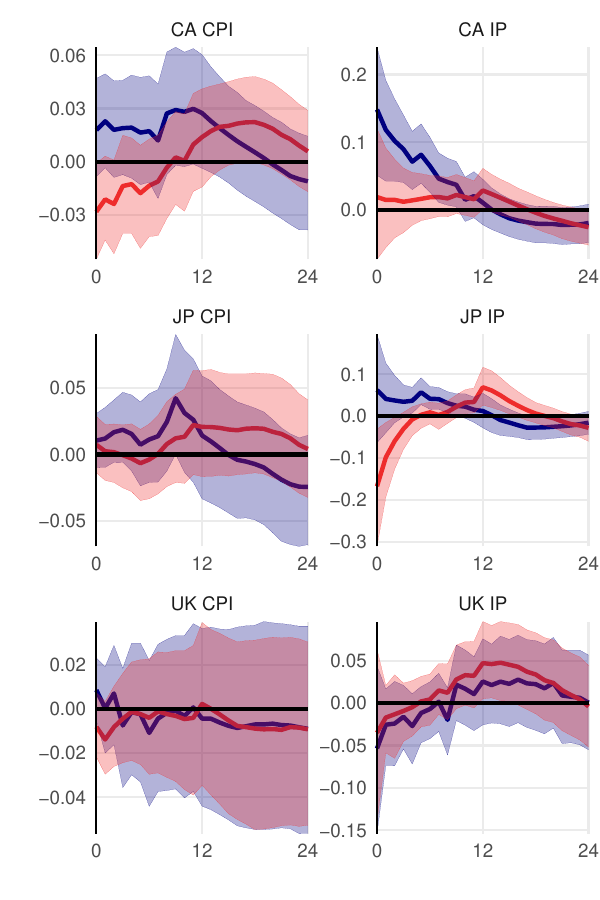}
    \end{subfigure}
    \caption{Spillovers to the indicated foreign country variables from a monetary policy and central bank information shock, respectively, in the EA (blue) and the US (red). Posterior median alongside the 68 percent credible set.}
    \label{fig:other-shock_domfor}
\end{figure}

Figure \ref{fig:other-shock_domfor} explores the spillover effects of monetary policy and central bank information shocks originating from the EA (blue) and the US (red) on selected foreign countries, namely Canada, the United Kingdom, and Japan. 

Panel (a) shows the results for a contractionary monetary policy shock, where EA monetary policy shocks tend to have a moderate and generally transient effect on CPI inflation and IP growth in foreign economies. The effect on UK consumer prices is most pronounced, reflecting its geographic proximity and economic interconnection to the EA, with a disinflationary effect for the first few months after impact. Similarly, US monetary policy shocks induce a stronger and more persistent response in CA consumer prices, while exhibiting less spillover effects to Japan or the UK.

Panel (b) illustrates the responses to central bank information shocks. Interestingly, EA shocks invoke a stronger response in consumer price inflation in CA and JP, while no significant response can be observed in the UK. In line with this result, the EA shock also raises IP growth in CA and JP, but it seems to have a limited spillover effect on the UK. Spillovers from US shocks have no significant effect on CPI inflation in the UK and JP, but show a negative effect on Canadian CPI inflation. In terms of industrial production, US shocks cause a decline in output growth in JP, whereas no significant effects can be observed in CA and the UK.

In summary, and in line with the results presented above, CI shocks tend to result in larger spillovers than monetary policy. The strength of spillovers from both an MP or CI shock tends to depend on the geographical closeness and the economic connectedness of the respective economies. It is worth mentioning that our analysis focuses on spillovers of EA and US shocks to the real economy of CA, JP and the UK and abstracts from spillovers to financial variables. Though surprising that US shocks do not have a larger effect on UK measures at first glance, this result is roughly in line with findings by \citet{cesa2022financial}.

Our findings highlight the need to reflect EA monetary policy decisions in global analyses, even though the US remains the driver of many worldwide economic developments. This insight corroborates findings by \citet{greenwood2021measuring}, who identify the EA as one of the world's most influential economies, and those by \citet{kearns2023explaining}, who find that the spillovers from EA monetary policy has gained importance over time.

\section{Closing remarks}\label{sec:conclusion}
This paper investigates the international spillovers of monetary policy and central bank information shocks between the US and the euro area (EA), using a novel identification approach that integrates external instruments, heteroskedasticity-based identification, and sign restrictions within a Bayesian PVAR model. Our framework offers a comprehensive way of modeling both static and dynamic spillovers, capturing the complex interdependencies between the EA, the US, and other major economies. This paper therefore contributes to the growing literature on cross-country monetary policy analysis and provides a robust tool for disentangling the effects of different types of shocks in a multicountry context.

The results reveal that both MP and CI shocks induce significant spillovers across economies, with CI shocks generally producing more pronounced effects, particularly in financial markets. The dominance of the US in the global economy is reflected in stronger and more persistent spillovers originating from US monetary policy, while EA shocks also exhibit noteworthy cross-border influence. The findings contribute to the understanding of dynamic interdependencies in global financial systems and highlight the asymmetric roles of the Fed and the ECB. The analysis underscores the importance of considering both MP and CI shocks in policymaking, as the latter can reveal market participants' expectations about the economic outlook. Moreover, the paper provides evidence that financial variables, such as stock markets and spreads, are more sensitive to cross-border shocks compared to real variables, such as industrial production and inflation.

{\small\setstretch{1}
\addcontentsline{toc}{section}{References}
\bibliographystyle{custom}
\bibliography{lit}}

\begin{thebibliography}{64}
\newcommand{\enquote}[1]{``#1''}
\providecommand{\natexlab}[1]{#1}

\bibitem[{Altavilla \emph{et~al.}(2019)Altavilla, Brugnolini, G{\"u}rkaynak,
  Motto, and Ragusa}]{altavilla2019measuring}
\textsc{Altavilla C, Brugnolini L, G{\"u}rkaynak RS, Motto R, and Ragusa G}
  (2019), \enquote{Measuring euro area monetary policy,} \emph{Journal of
  Monetary Economics} \textbf{108}, 162--179.

\bibitem[{Amir-Ahmadi \emph{et~al.}(2022)Amir-Ahmadi, Matthes, and
  Wang}]{amir2022understanding}
\textsc{Amir-Ahmadi P, Matthes C, and Wang MC} (2022), \enquote{Understanding
  Instruments in Macroeconomics--A Study of High-Frequency Identification,}
  Technical report, mimeo.

\bibitem[{Arias \emph{et~al.}(2018)Arias, Rubio-Ram{\'\i}rez, and
  Waggoner}]{arias2018inference}
\textsc{Arias JE, Rubio-Ram{\'\i}rez JF, and Waggoner DF} (2018),
  \enquote{Inference based on structural vector autoregressions identified with
  sign and zero restrictions: Theory and applications,} \emph{Econometrica}
  \textbf{86}(2), 685--720.

\bibitem[{Bai and Ng(2007)}]{bai2007determining}
\textsc{Bai J, and Ng S} (2007), \enquote{Determining the number of primitive
  shocks in factor models,} \emph{Journal of Business \& Economic Statistics}
  \textbf{25}(1), 52--60.

\bibitem[{Barigozzi \emph{et~al.}(2014)Barigozzi, Conti, and
  Luciani}]{doi:10.1111/obes.12038}
\textsc{Barigozzi M, Conti AM, and Luciani M} (2014), \enquote{Do Euro Area
  Countries Respond Asymmetrically to the Common Monetary Policy?} \emph{Oxford
  Bulletin of Economics and Statistics} \textbf{76}(5), 693--714.

\bibitem[{Bauer and Swanson(2023{\natexlab{a}})}]{bauer2023alternative}
\textsc{Bauer MD, and Swanson ET} (2023{\natexlab{a}}), \enquote{An Alternative
  Explanation for the "Fed Information Effect",} \emph{American Economic
  Review} \textbf{113}(3), 664--700.

\bibitem[{Bauer and Swanson(2023{\natexlab{b}})}]{bauer2023reassessment}
---{}---{}--- (2023{\natexlab{b}}), \enquote{A reassessment of monetary policy
  surprises and high-frequency identification,} \emph{NBER Macroeconomics
  Annual} \textbf{37}(1), 87--155.

\bibitem[{Baumeister and Hamilton(2015)}]{baumeister2015sign}
\textsc{Baumeister C, and Hamilton JD} (2015), \enquote{Sign restrictions,
  structural vector autoregressions, and useful prior information,}
  \emph{Econometrica} \textbf{83}(5), 1963--1999.

\bibitem[{Bertsche and Braun(2022)}]{bertsche2022identification}
\textsc{Bertsche D, and Braun R} (2022), \enquote{Identification of structural
  vector autoregressions by stochastic volatility,} \emph{Journal of Business
  \& Economic Statistics} \textbf{40}(1), 328--341.

\bibitem[{Borio and Hofmann(2017)}]{borio2017monetary}
\textsc{Borio CE, and Hofmann B} (2017), \enquote{Is monetary policy less
  effective when interest rates are persistently low?} \emph{BIS Working Paper}
  \textbf{628}.

\bibitem[{Breitenlechner \emph{et~al.}(2022)Breitenlechner, Georgiadis, and
  Schumann}]{breitenlechner2022goes}
\textsc{Breitenlechner M, Georgiadis G, and Schumann B} (2022), \enquote{What
  goes around comes around: How large are spillbacks from US monetary policy?}
  \emph{Journal of Monetary Economics} \textbf{131}, 45--60.

\bibitem[{Caldara and Herbst(2019)}]{caldara2019monetary}
\textsc{Caldara D, and Herbst E} (2019), \enquote{Monetary policy, real
  activity, and credit spreads: Evidence from Bayesian proxy SVARs,}
  \emph{American Economic Journal: Macroeconomics} \textbf{11}(1), 157--192.

\bibitem[{Carriero \emph{et~al.}(2024)Carriero, Marcellino, and
  Tornese}]{carriero2024blended}
\textsc{Carriero A, Marcellino M, and Tornese T} (2024), \enquote{Blended
  identification in structural VARs,} \emph{Journal of Monetary Economics}
  \textbf{146}, 103581.

\bibitem[{Carvalho \emph{et~al.}(2010)Carvalho, Polson, and
  Scott}]{carvalho2010horseshoe}
\textsc{Carvalho CM, Polson NG, and Scott JG} (2010), \enquote{The horseshoe
  estimator for sparse signals,} \emph{Biometrika} \textbf{97}(2), 465--480.

\bibitem[{Ca'Zorzi \emph{et~al.}(2023)Ca'Zorzi, Dedola, Georgiadis, Jarocinski,
  Stracca, and Strasser}]{ca2023making}
\textsc{Ca'Zorzi M, Dedola L, Georgiadis G, Jarocinski M, Stracca L, and
  Strasser G} (2023), \enquote{Making waves: Monetary policy and its asymmetric
  transmission in a globalized world,} \emph{International Journal of Central
  Banking} \textbf{19}(2), 95--144.

\bibitem[{Cesa-Bianchi and Sokol(2022)}]{cesa2022financial}
\textsc{Cesa-Bianchi A, and Sokol A} (2022), \enquote{Financial shocks, credit
  spreads, and the international credit channel,} \emph{Journal of
  International Economics} \textbf{135}, 103543.

\bibitem[{Chan \emph{et~al.}(2022)Chan, Eisenstat, and Yu}]{chan2022large}
\textsc{Chan J, Eisenstat E, and Yu X} (2022), \enquote{Large Bayesian VARs
  with factor stochastic volatility: Identification, order invariance and
  structural analysis,} \emph{arXiv} \textbf{2207.03988}.

\bibitem[{Ciccarelli \emph{et~al.}(2013)Ciccarelli, Maddaloni, and
  Peydr{\'o}}]{ciccarelli2013heterogeneous}
\textsc{Ciccarelli M, Maddaloni A, and Peydr{\'o} JL} (2013),
  \enquote{Heterogeneous transmission mechanism: monetary policy and financial
  fragility in the eurozone,} \emph{Economic Policy} \textbf{28}(75), 459--512.

\bibitem[{{Crespo Cuaresma} \emph{et~al.}(2019){Crespo Cuaresma}, Doppelhofer,
  Feldkircher, and Huber}]{crespo2019spillovers}
\textsc{{Crespo Cuaresma} J, Doppelhofer G, Feldkircher M, and Huber F} (2019),
  \enquote{Spillovers from US monetary policy: evidence from a time varying
  parameter global vector auto-regressive model,} \emph{Journal of the Royal
  Statistical Society: Series A (Statistics in Society)} \textbf{183}(3),
  831--861.

\bibitem[{{De Santis} and Surico(2013)}]{de2013bank}
\textsc{{De Santis} RA, and Surico P} (2013), \enquote{Bank lending and
  monetary transmission in the euro area,} \emph{Economic Policy}
  \textbf{28}(75), 423--457.

\bibitem[{Dedola \emph{et~al.}(2017)Dedola, Rivolta, and
  Stracca}]{dedola2017if}
\textsc{Dedola L, Rivolta G, and Stracca L} (2017), \enquote{If the {Fed}
  sneezes, who catches a cold?} \emph{Journal of International Economics}
  \textbf{108}, S23--S41.

\bibitem[{Degasperi \emph{et~al.}(2020)Degasperi, Hong, and
  Ricco}]{degasperi2020global}
\textsc{Degasperi R, Hong S, and Ricco G} (2020), \enquote{The global
  transmission of us monetary policy,} \emph{CEPR Discussion Paper}
  \textbf{14533}.

\bibitem[{Ehrmann and Fratzscher(2009)}]{ehrmann2009global}
\textsc{Ehrmann M, and Fratzscher M} (2009), \enquote{Global financial
  transmission of monetary policy shocks,} \emph{Oxford Bulletin of Economics
  and Statistics} \textbf{71}(6), 739--759.

\bibitem[{Ehrmann \emph{et~al.}(2011)Ehrmann, Fratzscher, and
  Rigobon}]{ehrmann2011stocks}
\textsc{Ehrmann M, Fratzscher M, and Rigobon R} (2011), \enquote{Stocks, bonds,
  money markets and exchange rates: measuring international financial
  transmission,} \emph{Journal of Applied Econometrics} \textbf{26}(6),
  948--974.

\bibitem[{Eickmeier and Ng(2015)}]{Eickmeier2011}
\textsc{Eickmeier S, and Ng T} (2015), \enquote{How do US credit supply shocks
  propagate internationally? A GVAR approach,} \emph{European Economic Review}
  \textbf{74}, 128--145.

\bibitem[{Feldkircher \emph{et~al.}(2020{\natexlab{a}})Feldkircher, Gruber, and
  Huber}]{feldkircher2020international}
\textsc{Feldkircher M, Gruber T, and Huber F} (2020{\natexlab{a}}),
  \enquote{International effects of a compression of euro area yield curves,}
  \emph{Journal of Banking \& Finance} \textbf{113}, 105533.

\bibitem[{Feldkircher \emph{et~al.}(2022)Feldkircher, Huber, Koop, and
  Pfarrhofer}]{feldkircher2022approximate}
\textsc{Feldkircher M, Huber F, Koop G, and Pfarrhofer M} (2022),
  \enquote{Approximate Bayesian inference and forecasting in huge-dimensional
  multicountry VARs,} \emph{International Economic Review} \textbf{63}(4),
  1625--1658.

\bibitem[{Feldkircher \emph{et~al.}(2020{\natexlab{b}})Feldkircher, Huber, and
  Pfarrhofer}]{feldkircher2020factor}
\textsc{Feldkircher M, Huber F, and Pfarrhofer M} (2020{\natexlab{b}}),
  \enquote{Factor Augmented Vector Autoregressions, Panel VARs, and Global
  VARs,} in \textsc{P~Fuleky} (ed.) \enquote{Macroeconomic Forecasting in the
  Era of Big Data,} 65--93, Springer.

\bibitem[{Gambetti \emph{et~al.}(2023)Gambetti, Korobilis, Tsoukalas, and
  Zanetti}]{gambetti2023agreed}
\textsc{Gambetti L, Korobilis D, Tsoukalas J, and Zanetti F} (2023),
  \enquote{Agreed and disagreed uncertainty,} \emph{arXiv} \textbf{2302.01621}.

\bibitem[{Gardner \emph{et~al.}(2022)Gardner, Scotti, and
  Vega}]{gardner2022words}
\textsc{Gardner B, Scotti C, and Vega C} (2022), \enquote{Words speak as loudly
  as actions: Central bank communication and the response of equity prices to
  macroeconomic announcements,} \emph{Journal of Econometrics} \textbf{231}(2),
  387--409.

\bibitem[{Georgiadis(2015)}]{georgiadis2015examining}
\textsc{Georgiadis G} (2015), \enquote{Examining asymmetries in the
  transmission of monetary policy in the euro area: Evidence from a mixed
  cross-section global VAR model,} \emph{European Economic Review} \textbf{75},
  195--215.

\bibitem[{Georgiadis and Jaroci\'nski(2023)}]{georgiadis2023global}
\textsc{Georgiadis G, and Jaroci\'nski M} (2023), \enquote{Global spillovers
  from multi-dimensional US monetary policy,} \emph{ECB Working Paper}
  \textbf{2881}.

\bibitem[{Gerko and Rey(2017)}]{gerko2017monetary}
\textsc{Gerko E, and Rey H} (2017), \enquote{Monetary policy in the capitals of
  capital,} \emph{Journal of the European Economic Association} \textbf{15}(4),
  721--745.

\bibitem[{Gilchrist \emph{et~al.}(2009)Gilchrist, Yankov, and
  Zakraj{\v{s}}ek}]{gilchrist2009credit}
\textsc{Gilchrist S, Yankov V, and Zakraj{\v{s}}ek E} (2009), \enquote{Credit
  market shocks and economic fluctuations: Evidence from corporate bond and
  stock markets,} \emph{Journal of Monetary Economics} \textbf{56}(4),
  471--493.

\bibitem[{Gilchrist and Zakraj{\v{s}}ek(2012)}]{gilchrist2012credit}
\textsc{Gilchrist S, and Zakraj{\v{s}}ek E} (2012), \enquote{Credit spreads and
  business cycle fluctuations,} \emph{American economic review}
  \textbf{102}(4), 1692--1720.

\bibitem[{Greenwood-Nimmo \emph{et~al.}(2021)Greenwood-Nimmo, Nguyen, and
  Shin}]{greenwood2021measuring}
\textsc{Greenwood-Nimmo M, Nguyen VH, and Shin Y} (2021), \enquote{Measuring
  the connectedness of the global economy,} \emph{International Journal of
  Forecasting} \textbf{37}(2), 899--919.

\bibitem[{Griller \emph{et~al.}(2024)Griller, Huber, and
  Pfarrhofer}]{griller2024financial}
\textsc{Griller S, Huber F, and Pfarrhofer M} (2024), \enquote{Financial
  markets and legal challenges to unconventional monetary policy,}
  \emph{European Economic Review} \textbf{163}, 104680.

\bibitem[{G{\"u}rkaynak \emph{et~al.}(2020)G{\"u}rkaynak,
  K{\i}sac{\i}ko{\u{g}}lu, and Wright}]{gurkaynak2020missing}
\textsc{G{\"u}rkaynak RS, K{\i}sac{\i}ko{\u{g}}lu B, and Wright JH} (2020),
  \enquote{Missing events in event studies: Identifying the effects of
  partially measured news surprises,} \emph{American Economic Review}
  \textbf{110}(12), 3871--3912.

\bibitem[{G\"{u}rkaynak \emph{et~al.}(2005)G\"{u}rkaynak, Sack, and
  Swanson}]{gurkaynak2005actions}
\textsc{G\"{u}rkaynak RS, Sack BP, and Swanson ET} (2005), \enquote{Do actions
  speak louder than words? The response of asset prices to monetary policy
  actions and statements,} \emph{International Journal of Central Banking}
  \textbf{1}(1).

\bibitem[{Hausman and Wongswan(2011)}]{hausman2011global}
\textsc{Hausman J, and Wongswan J} (2011), \enquote{Global asset prices and
  FOMC announcements,} \emph{Journal of International Money and Finance}
  \textbf{30}(3), 547--571.

\bibitem[{Huber \emph{et~al.}(2023)Huber, Krisztin, and
  Pfarrhofer}]{huber2018bayesian}
\textsc{Huber F, Krisztin T, and Pfarrhofer M} (2023), \enquote{A Bayesian
  panel vector autoregression to analyze the impact of climate shocks on
  high-income economies,} \emph{Annals of Applied Statistics} \textbf{17}(2),
  1543--1573.

\bibitem[{Huber \emph{et~al.}(2024)Huber, Matthes, and
  Pfarrhofer}]{huber2024general}
\textsc{Huber F, Matthes C, and Pfarrhofer M} (2024), \enquote{General
  Seemingly Unrelated Local Projections,} \emph{arXiv} \textbf{2410.17105}.

\bibitem[{Jaroci{\'n}ski(2022)}]{jarocinski2022central}
\textsc{Jaroci{\'n}ski M} (2022), \enquote{Central bank information effects and
  transatlantic spillovers,} \emph{Journal of International Economics}
  \textbf{139}, 103683.

\bibitem[{Jaroci{\'n}ski and Karadi(2020)}]{jarocinski2020deconstructing}
\textsc{Jaroci{\'n}ski M, and Karadi P} (2020), \enquote{Deconstructing
  monetary policy surprises -- the role of information shocks,} \emph{American
  Economic Journal: Macroeconomics} \textbf{12}(2), 1--43.

\bibitem[{Kastner and Huber(2020)}]{kastner2020sparse}
\textsc{Kastner G, and Huber F} (2020), \enquote{Sparse Bayesian vector
  autoregressions in huge dimensions,} \emph{Journal of Forecasting}
  \textbf{39}(7), 1142--1165.

\bibitem[{Kearns \emph{et~al.}(2023)Kearns, Schrimpf, and
  Xia}]{kearns2023explaining}
\textsc{Kearns J, Schrimpf A, and Xia FD} (2023), \enquote{Explaining monetary
  spillovers: The matrix reloaded,} \emph{Journal of Money, Credit and Banking}
  \textbf{55}(6), 1535--1568.

\bibitem[{Kim(2001)}]{kim2001international}
\textsc{Kim S} (2001), \enquote{International transmission of US monetary
  policy shocks: Evidence from VAR's,} \emph{Journal of Monetary Economics}
  \textbf{48}(2), 339--372.

\bibitem[{Kim \emph{et~al.}(1998)Kim, Shephard, and Chib}]{kim1998stochastic}
\textsc{Kim S, Shephard N, and Chib S} (1998), \enquote{Stochastic volatility:
  likelihood inference and comparison with ARCH models,} \emph{Review of
  Economic Studies} \textbf{65}(3), 361--393.

\bibitem[{Korobilis(2022)}]{korobilis2022new}
\textsc{Korobilis D} (2022), \enquote{A new algorithm for structural
  restrictions in Bayesian vector autoregressions,} \emph{European Economic
  Review} \textbf{148}, 104241.

\bibitem[{Lewis(2021)}]{lewis2021identifying}
\textsc{Lewis DJ} (2021), \enquote{Identifying shocks via time-varying
  volatility,} \emph{The Review of Economic Studies} \textbf{88}(6),
  3086--3124.

\bibitem[{Li \emph{et~al.}(2024)Li, Plagborg-M{\o}ller, and Wolf}]{li2024local}
\textsc{Li D, Plagborg-M{\o}ller M, and Wolf CK} (2024), \enquote{Local
  projections vs. VARs: Lessons from thousands of DGPs,} \emph{Journal of
  Econometrics} 105722.

\bibitem[{Makalic and Schmidt(2015)}]{makalic2015simple}
\textsc{Makalic E, and Schmidt DF} (2015), \enquote{A simple sampler for the
  horseshoe estimator,} \emph{IEEE Signal Processing Letters} \textbf{23}(1),
  179--182.

\bibitem[{McCracken and Ng(2016)}]{mccracken2016fred}
\textsc{McCracken M, and Ng S} (2016), \enquote{FRED-MD: A monthly database for
  macroeconomic research,} \emph{Journal of Business \& Economic Statistics}
  \textbf{34}(4), 574--589.

\bibitem[{Mertens and Ravn(2013)}]{mertens2013dynamic}
\textsc{Mertens K, and Ravn MO} (2013), \enquote{The dynamic effects of
  personal and corporate income tax changes in the United States,}
  \emph{American Economic Review} \textbf{103}(4), 1212--1247.

\bibitem[{Miranda-Agrippino and Nenova(2022)}]{miranda2022tale}
\textsc{Miranda-Agrippino S, and Nenova T} (2022), \enquote{A tale of two
  global monetary policies,} \emph{Journal of International Economics}
  \textbf{136}, 103606.

\bibitem[{Miranda-Agrippino and Rey(2020)}]{miranda2020us}
\textsc{Miranda-Agrippino S, and Rey H} (2020), \enquote{US monetary policy and
  the global financial cycle,} \emph{The Review of Economic Studies}
  \textbf{87}(6), 2754--2776.

\bibitem[{Nakamura and Steinsson(2018)}]{nakamura2018high}
\textsc{Nakamura E, and Steinsson J} (2018), \enquote{High-frequency
  identification of monetary non-neutrality: the information effect,}
  \emph{Quarterly Journal of Economics} \textbf{133}(3), 1283--1330.

\bibitem[{Plagborg-M{\o}ller and Wolf(2021)}]{plagborg2021local}
\textsc{Plagborg-M{\o}ller M, and Wolf CK} (2021), \enquote{Local projections
  and VARs estimate the same impulse responses,} \emph{Econometrica}
  \textbf{89}(2), 955--980.

\bibitem[{Potjagailo(2017)}]{potjagailo2017spillover}
\textsc{Potjagailo G} (2017), \enquote{Spillover effects from Euro area
  monetary policy across Europe: A factor-augmented VAR approach,}
  \emph{Journal of International Money and Finance} \textbf{72}, 127--147.

\bibitem[{Ramey(2016)}]{ramey2016macroeconomic}
\textsc{Ramey VA} (2016), \enquote{Macroeconomic shocks and their propagation,}
  \emph{Handbook of Macroeconomics} \textbf{2}, 71--162.

\bibitem[{Rigobon(2003)}]{rigobon2003identification}
\textsc{Rigobon R} (2003), \enquote{Identification through heteroskedasticity,}
  \emph{Review of Economics and Statistics} \textbf{85}(4), 777--792.

\bibitem[{Schlaak \emph{et~al.}(2023)Schlaak, Rieth, and
  Podstawski}]{schlaak2023monetary}
\textsc{Schlaak T, Rieth M, and Podstawski M} (2023), \enquote{Monetary policy,
  external instruments, and heteroskedasticity,} \emph{Quantitative Economics}
  \textbf{14}(1), 161--200.

\bibitem[{Sentana and Fiorentini(2001)}]{sentana2001identification}
\textsc{Sentana E, and Fiorentini G} (2001), \enquote{Identification,
  estimation and testing of conditionally heteroskedastic factor models,}
  \emph{Journal of Econometrics} \textbf{102}(2), 143--164.

\bibitem[{Tenreyro and Thwaites(2016)}]{tenreyro2016}
\textsc{Tenreyro S, and Thwaites G} (2016), \enquote{Pushing on a String: US
  Monetary Policy Is Less Powerful in Recessions,} \emph{American Economic
  Journal: Macroeconomics} \textbf{8}(4), 43--74.

\end{thebibliography}

\newpage

\begin{appendices}\crefalias{section}{appsec}
\begin{center}
\LARGE{\sffamily\textbf{Appendix}}
\end{center}

\setcounter{equation}{0}
\setcounter{figure}{0}
\renewcommand\theequation{A.\arabic{equation}}
\section{Data}\label{app:data}
We use a panel of monthly data ranging from 1999:01 to 2019:12 for the Euro area [EA], Canada [CA], Japan [JP], the United Kingdom [UK], and the United States [US].

\begin{table*}[h]
\caption{Data, transformations and source.}\label{tab:appdata}\vspace*{-1.5em}
\begin{center}
\begin{threeparttable}
\footnotesize
\begin{tabular*}{\textwidth}{@{\extracolsep{\fill}} llll}
  \toprule
\textbf{Variable} & \textbf{Details} & \textbf{Transformation} & \textbf{Source} \\
\midrule
\multicolumn{4}{l}{\textit{Euro Area}} \\
\midrule
Surprises in interest rates & --- & normalized & \citet{jarocinski2020deconstructing} \\
Surprises in Euro Stoxx 50 & --- & normalized & \citet{jarocinski2020deconstructing} \\
Interest rate & 1-year Bund bond yield & BPs & Macrobond \\
Stock market & Euro Stoxx 50 index & 100 log-diff & ECB SDW \\
Industrial production & -- & y-o-y & OECD \\
Consumer price index & -- & y-o-y & OECD/FRED \\
Financial conditions & OAS & BPs & FRED \\
\midrule
\multicolumn{4}{l}{\textit{United States}} \\
\midrule
Suprises in Fed funds futures & --- & normalized & \citet{jarocinski2020deconstructing} \\
Suprises in S\&P500 & --- & normalized & \citet{jarocinski2020deconstructing} \\
Interest rate & 1-year US bond yield & BPs & Macrobond \\
Stock market & S\&P 500 index & 100 log-diff & FRED-MD \\
Industrial production & INDPRO & y-o-y & FRED-MD\\
Consumer price index & CPIAUCSL & y-o-y & FRED-MD \\
Financial conditions & Excess bond premium & BPs & \citet{gilchrist2009credit} \\
\midrule
\multicolumn{4}{l}{\textit{International variables}} \\
\midrule
Oil prices & OILPRICEx & log & FRED-MD\\
USD/EUR exchange rate & USDEUR & log & FRED-MD \\
 \midrule
\multicolumn{4}{l}{\textit{Country-specific}} \\
\midrule
Consumer price index & *CPIALLMINMEI & y-o-y & OECD/FRED \\
Industrial production & MEI\_REAL PRINTO01 & y-o-y & OECD/FRED \\
\bottomrule
\end{tabular*}
\begin{tablenotes}[para,flushleft]
\scriptsize{\textit{Notes}: Data from the FRED database of the Federal Reserve Bank of St. Louis is available for download at \href{https://fred.stlouisfed.org/}{fred.stlouisfed.org}. Data from FRED-MD, a monthly database for macroeconomic research \citep{mccracken2016fred}, are retrieved from \href{https://research.stlouisfed.org/econ/mccracken/fred-databases/}{fred-databases}.
Asterisks ($\ast$) in detailed data names refer to country codes. \citet{jarocinski2020deconstructing} refers to updated shock series downloaded from Marek Jaroci\'{n}ski's \href{https://marekjarocinski.github.io/jkshocks/jkshocks.html}{webpage}. Abbreviations: year-on-year differences (y-o-y).}
\end{tablenotes}
\end{threeparttable}
\end{center}
\end{table*}

\clearpage
\section{Additional Empirical Results}
\begin{figure}[ht]
    \begin{subfigure}[t]{\textwidth}
    \caption{Euro area (EA)}
    \includegraphics[width=\textwidth]{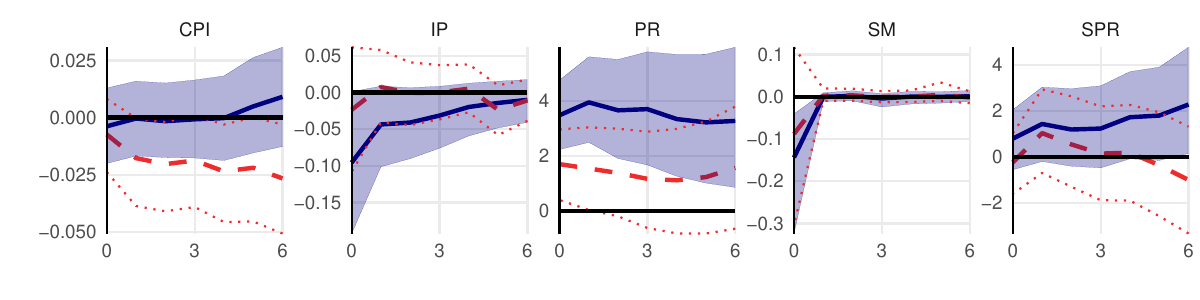}
    \end{subfigure}
    \begin{subfigure}[t]{\textwidth}
    \caption{United States (US)}
    \includegraphics[width=\textwidth]{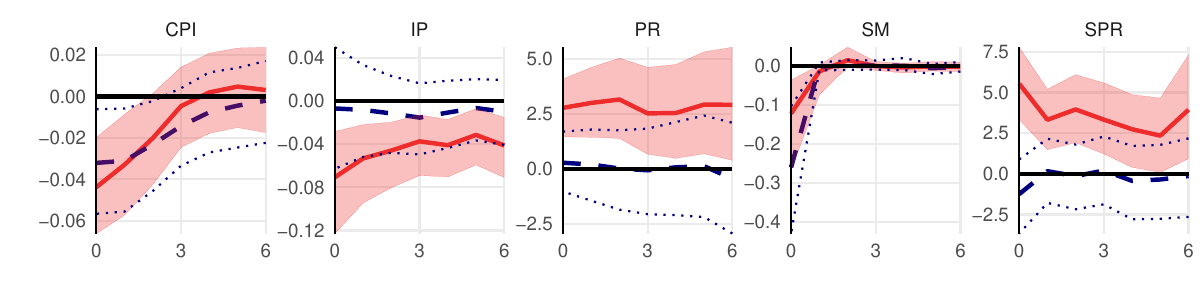}
    \end{subfigure}
    \caption{Responses of the indicated variables to a monetary policy shock originating in the EA (blue) and the US (red) for the first six months after impact. The response of each variable to the respective domestic shock is in solid colors, the response to the foreign shock is indicated by the dashed/dotted lines. Posterior median alongside the 68 percent credible set.}
    \label{fig:mp-shock_domfor_zoom}
\end{figure}

\begin{figure}[t]
    \begin{subfigure}[t]{\textwidth}
    \caption{Euro area (EA)}
    \includegraphics[width=\textwidth]{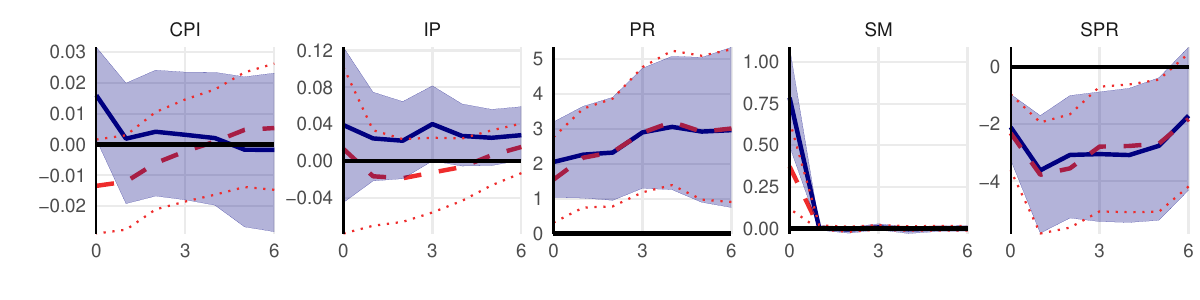}
    \end{subfigure}
    \begin{subfigure}[t]{\textwidth}
    \caption{United States (US)}
    \includegraphics[width=\textwidth]{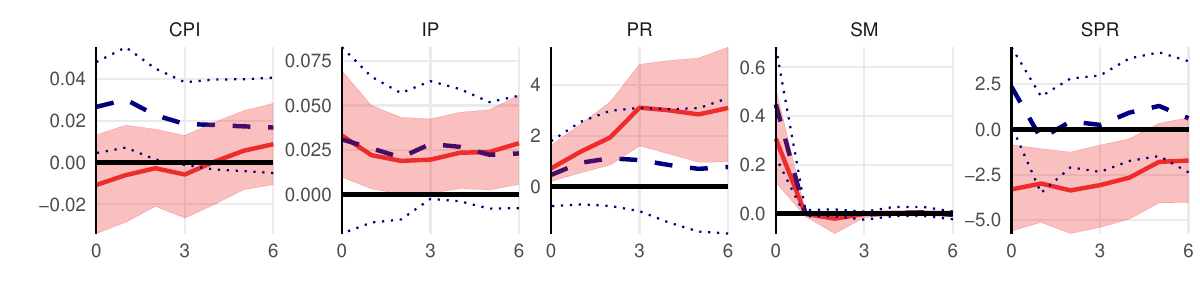}
    \end{subfigure}
    \caption{Responses for the indicated variables to a central bank information shock originating in the US (red) and the EA (blue) for the first six months after impact. The response of each variable to the respective domestic shock is in solid colors, the response to the foreign shock is indicated by the dashed/dotted lines. Posterior median alongside the 68 percent credible set.}
    \label{fig:ci-shock_domfor_zoom}
\end{figure}

\end{appendices}

\end{document}